\documentclass[%
 reprint,
 amsmath,amssymb,
 aps,
 prmaterials,
]{revtex4-2}

\usepackage{graphicx}
\usepackage{dcolumn}
\usepackage{bm}
\usepackage{hyperref}
\usepackage{datatool} 
\usepackage{pgffor} 
\usepackage{float} 
\usepackage{subcaption} 
\newcommand{\unit}[2]{${#1 \; \mathrm{#2}}$}

\usepackage{datatool} 
\usepackage{pgffor} 

\DTLloaddb[keys={Elem,Def,Ef,P11,P22,P33,P12,P23,P31,O11,O22,O33,O12,O23,O31,RelVol},
headers={\shortstack{},
\shortstack{Defect name},\shortstack{$E_f$ (eV)},
\shortstack{$P_{11}$ (eV)},\shortstack{$P_{22}$},\shortstack{$P_{33}$},
\shortstack{$P_{12}$},\shortstack{$P_{23}$},\shortstack{$P_{31}$},
\shortstack{$\Omega_{11}$},\shortstack{$\Omega_{22}$},\shortstack{$\Omega_{33}$},
\shortstack{$\Omega_{12}$},\shortstack{$\Omega_{23}$},\shortstack{$\Omega_{31}$},
\shortstack{$\Omega_{rel}$}
}]{defect_data}{data/defect_data.txt}

\newcommand{\DefectDataTable}[4]{%
\setlength\tabcolsep{2pt}
\begin{table*}
\caption{#1 (#3) point defect dipole tensor data computed using the GGA-#4 functional. Formation energy $E_f$  is given in eV units, dipole tensor components $P_{ij}$ also in eV units, and relaxation volume tensor components $\Omega_{ij}$ and its trace $\Omega_{rel}=\Omega_{ii}$ are given in atomic volume units.}
\centering
\begin{ruledtabular}
\begin{tabular}{*{15}{c}}
 #3\_#4 &
 $E_f$ &
 $P_{11}$ &  $P_{22}$ &  $P_{33}$ &
 $P_{12}$ &  $P_{23}$ &  $P_{31}$ &
 $\Omega_{11}$ &  $\Omega_{22}$ &  $\Omega_{33}$ &
 $\Omega_{12}$ &  $\Omega_{23}$ &  $\Omega_{31}$ &
 $\Omega_{rel}$ %
\\ \hline
\DTLforeach*[\DTLiseq{\Elem}{#2}]{defect_data}{%
\Elem=Elem,\Def=Def,\Ef=Ef, \Pa=P11, \Pb=P22, \Pc=P33, \Pd=P12, \Pe=P23, \Pf =P31, \Oa=O11, \Ob=O22, \Oc=O33, \Od=O12, \Oe=O23, \Of =O31, \RV=RelVol}{%
\\
\Def & 
\Ef &
\Pa & \Pb & \Pc & \Pd & \Pe & \Pf &
\Oa & \Ob & \Oc & \Od & \Oe & \Of & \RV}
\end{tabular}
\end{ruledtabular}
\end{table*}%
}

\begin{document}
\begin{filecontents*}{data/defect_data.txt}
\# Elem , Defect  , E_f (eV), P11 (eV), P22, P33, P12, P23, P31, O11 (AtVol), O22, O33, O12, O23, O31, Rel Vol 
Be   , Vac  ,    0.610,   -4.004,   -4.004,    3.684,    0.000,    0.000,    0.000,   -0.239,   -0.239,    0.201,    0.000,    0.000,    0.000,   -0.277
Be   , BO   ,    3.900,   18.692,   18.692,   -5.766,    0.000,    0.000,    0.000,    1.103,    1.103,   -0.345,    0.000,    0.000,    0.000,    1.861
Be   , BS   ,    4.210,   19.214,   18.008,   -5.331,    0.000,    0.000,    0.000,    1.161,    1.035,   -0.322,    0.000,    0.000,    0.000,    1.873
Be   , C    ,    4.330,   21.923,   12.302,    0.273,    5.402,    8.208,    9.356,    1.505,    0.501,   -0.027,    1.127,    0.978,    1.115,    1.979
Be   , O    ,    5.130,   16.508,   16.508,    1.730,    0.000,    0.000,    0.000,    0.966,    0.966,    0.050,    0.000,    0.000,    0.000,    1.981
Be   , S    ,    5.280,    7.500,    7.500,   23.420,    0.000,    0.000,    0.000,    0.411,    0.411,    1.199,    0.000,    0.000,    0.000,    2.021
Be_PBEsol, Vac  ,    0.580,   -4.787,   -4.787,    5.004,    0.000,    0.000,    0.000,   -0.277,   -0.277,    0.260,    0.000,    0.000,    0.000,   -0.294
Be_PBEsol, BO   ,    4.250,   15.675,   15.675,   -1.025,    0.000,    0.000,    0.000,    0.901,    0.901,   -0.065,    0.000,    0.000,    0.000,    1.736
Be_PBEsol, BS   ,    4.350,   16.399,   16.441,   -1.371,    0.000,    0.000,    0.000,    0.942,    0.945,   -0.084,    0.000,    0.000,    0.000,    1.803
Be_PBEsol, C    ,    4.470,   18.945,   11.035,    4.037,    4.742,    6.850,    8.214,    1.104,    0.615,    0.194,    0.586,    0.852,    1.022,    1.912
Be_PBEsol, S    ,    5.340,    3.666,    3.666,   30.178,    0.000,    0.000,    0.000,    0.198,    0.198,    1.539,    0.000,    0.000,    0.000,    1.935
Be_PBEsol, O    ,    5.500,   14.647,   14.647,    4.058,    0.000,    0.000,    0.000,    0.839,    0.839,    0.195,    0.000,    0.000,    0.000,    1.874
Mg   , Vac  ,    0.730,   -1.869,   -1.869,   -1.290,    0.000,    0.000,    0.000,   -0.133,   -0.133,   -0.053,    0.000,    0.000,    0.000,   -0.319
Mg   , C    ,    2.160,   10.891,    8.752,    8.332,    0.566,    1.853,    0.980,    0.832,    0.485,    0.459,    0.183,    0.632,    0.334,    1.776
Mg   , S    ,    2.250,    9.085,    9.085,    9.946,    0.000,    0.000,    0.000,    0.551,    0.551,    0.685,    0.000,    0.000,    0.000,    1.787
Mg   , BO   ,    2.260,   12.178,   12.178,    2.695,    0.000,    0.000,    0.000,    1.016,    1.016,   -0.321,    0.000,    0.000,    0.000,    1.711
Mg   , O    ,    2.290,    8.737,    8.737,   10.935,    0.000,    0.000,    0.000,    0.494,    0.494,    0.819,    0.000,    0.000,    0.000,    1.807
Mg   , BS   ,    2.340,   15.305,    9.885,    2.418,    0.000,    0.000,    0.000,    1.499,    0.621,   -0.375,    0.000,    0.000,    0.000,    1.746
Sc   , Vac  ,    1.910,   -2.543,   -2.543,   -1.761,    0.000,    0.000,    0.000,   -0.109,   -0.109,   -0.053,    0.000,    0.000,    0.000,   -0.271
Sc   , BO   ,    2.020,   10.059,   10.059,    5.910,    0.000,    0.000,    0.000,    0.446,    0.446,    0.133,    0.000,    0.000,    0.000,    1.025
Sc   , O    ,    2.390,    9.102,    9.102,    8.233,    0.000,    0.000,    0.000,    0.362,    0.362,    0.327,    0.000,    0.000,    0.000,    1.051
Sc   , BS   ,    2.540,   11.615,    9.543,    7.533,    0.000,    0.000,    0.000,    0.558,    0.342,    0.235,    0.000,    0.000,    0.000,    1.135
Sc   , C    ,    2.710,    7.081,    6.969,   13.423,    0.724,    0.097,    1.255,    0.183,    0.172,    0.759,    0.150,    0.022,    0.280,    1.114
Sc   , S    ,    2.920,    4.125,    4.125,   16.727,    0.000,    0.000,    0.000,   -0.023,   -0.023,    1.080,    0.000,    0.000,    0.000,    1.033
Y    , Vac  ,    1.860,   -3.247,   -3.247,   -2.041,    0.000,    0.000,    0.000,   -0.149,   -0.149,   -0.045,    0.000,    0.000,    0.000,   -0.342
Y    , BO   ,    2.080,    9.379,    9.379,    5.632,    0.000,    0.000,    0.000,    0.433,    0.433,    0.110,    0.000,    0.000,    0.000,    0.977
Y    , O    ,    2.520,    8.432,    8.432,    7.859,    0.000,    0.000,    0.000,    0.348,    0.348,    0.294,    0.000,    0.000,    0.000,    0.989
Y    , BS   ,    2.540,   10.827,    8.130,    6.514,    0.000,    0.000,    0.000,    0.553,    0.299,    0.169,    0.000,    0.000,    0.000,    1.020
Y    , C    ,    2.790,    6.482,    6.061,   12.364,    1.148,    0.364,    1.989,    0.180,    0.140,    0.674,    0.216,    0.071,    0.388,    0.994
Y    , S    ,    2.930,    2.994,    2.994,   15.058,    0.000,    0.000,    0.000,   -0.061,   -0.061,    0.961,    0.000,    0.000,    0.000,    0.838
Lu   , Vac  ,    1.880,   -2.765,   -2.765,   -1.863,    0.000,    0.000,    0.000,   -0.116,   -0.116,   -0.057,    0.000,    0.000,    0.000,   -0.289
Lu   , BO   ,    2.620,   12.130,   12.130,    6.094,    0.000,    0.000,    0.000,    0.535,    0.535,    0.113,    0.000,    0.000,    0.000,    1.183
Lu   , O    ,    3.090,   10.255,   10.255,   10.617,    0.000,    0.000,    0.000,    0.389,    0.389,    0.450,    0.000,    0.000,    0.000,    1.227
Lu   , BS   ,    3.100,   14.979,   10.097,    7.139,    0.000,    0.000,    0.000,    0.758,    0.328,    0.172,    0.000,    0.000,    0.000,    1.258
Lu   , S    ,    3.600,    5.830,    5.830,   17.760,    0.000,    0.000,    0.000,    0.085,    0.085,    1.015,    0.000,    0.000,    0.000,    1.185
Ti   , Vac  ,    2.030,   -3.933,   -3.933,   -6.786,    0.000,    0.000,    0.000,   -0.056,   -0.056,   -0.293,    0.000,    0.000,    0.000,   -0.404
Ti   , BO   ,    2.380,   13.722,   13.722,   10.015,    0.000,    0.000,    0.000,    0.458,    0.458,    0.126,    0.000,    0.000,    0.000,    1.042
Ti   , O    ,    2.520,   13.816,   13.816,   11.418,    0.000,    0.000,    0.000,    0.435,    0.435,    0.215,    0.000,    0.000,    0.000,    1.085
Ti   , BS   ,    2.600,   10.426,   18.336,    9.335,    0.000,    0.000,    0.000,    0.065,    0.939,    0.057,    0.000,    0.000,    0.000,    1.060
Ti   , S    ,    2.810,   10.169,   10.169,   18.430,    0.000,    0.000,    0.000,    0.127,    0.127,    0.814,    0.000,    0.000,    0.000,    1.068
Zr   , Vac  ,    1.940,   -5.156,   -5.156,   -8.022,    0.000,    0.000,    0.000,   -0.085,   -0.085,   -0.267,    0.000,    0.000,    0.000,   -0.437
Zr   , BO   ,    2.690,   16.694,   16.694,   10.344,    0.000,    0.000,    0.000,    0.551,    0.551,   -0.009,    0.000,    0.000,    0.000,    1.092
Zr   , BS   ,    2.870,   13.627,   21.842,    9.409,    0.000,    0.000,    0.000,    0.229,    0.997,   -0.098,    0.000,    0.000,    0.000,    1.128
Zr   , O    ,    2.910,   14.795,   14.795,   16.624,    0.000,    0.000,    0.000,    0.357,    0.357,    0.408,    0.000,    0.000,    0.000,    1.122
Zr   , S    ,    3.050,   11.084,   11.084,   21.505,    0.000,    0.000,    0.000,    0.108,    0.108,    0.811,    0.000,    0.000,    0.000,    1.028
Zr   , C    ,    3.320,   10.049,    2.706,   22.461,    0.478,    6.359,    0.828,    0.228,   -0.458,    1.030,    0.089,    1.695,    0.221,    0.800
Hf   , Vac  ,    2.220,   -5.529,   -5.529,   -7.159,    0.000,    0.000,    0.000,   -0.105,   -0.105,   -0.186,    0.000,    0.000,    0.000,   -0.397
Hf   , BO   ,    3.970,   20.379,   20.379,   12.701,    0.000,    0.000,    0.000,    0.550,    0.550,    0.081,    0.000,    0.000,    0.000,    1.181
Hf   , BS   ,    4.260,   17.488,   25.202,   11.591,    0.000,    0.000,    0.000,    0.347,    0.846,    0.009,    0.000,    0.000,    0.000,    1.202
Hf   , O    ,    4.420,   17.554,   17.554,   21.859,    0.000,    0.000,    0.000,    0.345,    0.345,    0.553,    0.000,    0.000,    0.000,    1.243
Hf   , S    ,    4.620,   12.616,   12.616,   30.220,    0.000,    0.000,    0.000,    0.077,    0.077,    1.040,    0.000,    0.000,    0.000,    1.193
Tc   , Vac  ,    2.620,   -9.049,   -9.049,   -9.326,    0.000,    0.000,    0.000,   -0.111,   -0.111,   -0.116,    0.000,    0.000,    0.000,   -0.338
Tc   , S    ,    4.390,   48.246,   48.246,   46.296,    0.000,    0.000,    0.000,    0.612,    0.612,    0.535,    0.000,    0.000,    0.000,    1.760
Tc   , BO   ,    5.200,   52.274,   52.274,   26.679,    0.000,    0.000,    0.000,    0.816,    0.816,   -0.009,    0.000,    0.000,    0.000,    1.623
Tc   , BS   ,    5.560,   57.215,   46.418,   24.881,    0.000,    0.000,    0.000,    1.042,    0.595,   -0.048,    0.000,    0.000,    0.000,    1.590
Tc   , O    ,    5.880,   37.384,   37.384,   64.750,    0.000,    0.000,    0.000,    0.287,    0.287,    1.139,    0.000,    0.000,    0.000,    1.713
Re   , Vac  ,    3.200,   -9.555,   -9.555,  -12.995,    0.000,    0.000,    0.000,   -0.076,   -0.076,   -0.158,    0.000,    0.000,    0.000,   -0.310
Re   , S    ,    6.950,   69.410,   69.410,   59.367,    0.000,    0.000,    0.000,    0.721,    0.721,    0.473,    0.000,    0.000,    0.000,    1.915
Re   , BO   ,    7.840,   74.735,   74.735,   33.057,    0.000,    0.000,    0.000,    0.926,    0.926,   -0.085,    0.000,    0.000,    0.000,    1.766
Re   , BS   ,    8.340,   92.354,   56.586,   27.551,    0.000,    0.000,    0.000,    1.529,    0.368,   -0.189,    0.000,    0.000,    0.000,    1.708
Re   , O    ,    8.460,   53.148,   53.148,   86.902,    0.000,    0.000,    0.000,    0.352,    0.352,    1.160,    0.000,    0.000,    0.000,    1.864
Ru   , Vac  ,    2.620,   -7.414,   -7.414,  -10.303,    0.000,    0.000,    0.000,   -0.082,   -0.082,   -0.144,    0.000,    0.000,    0.000,   -0.309
Ru   , S    ,    8.430,   50.118,   50.118,   62.809,    0.000,    0.000,    0.000,    0.591,    0.591,    0.831,    0.000,    0.000,    0.000,    2.012
Ru   , BO   ,    8.480,   57.347,   57.347,   36.119,    0.000,    0.000,    0.000,    0.853,    0.853,    0.194,    0.000,    0.000,    0.000,    1.901
Ru   , O    ,    8.520,   49.671,   49.671,   55.008,    0.000,    0.000,    0.000,    0.621,    0.621,    0.670,    0.000,    0.000,    0.000,    1.913
Ru   , BS   ,    9.150,   65.662,   46.200,   41.224,    0.000,    0.000,    0.000,    1.091,    0.514,    0.316,    0.000,    0.000,    0.000,    1.921
Os   , Vac  ,    2.980,   -9.967,   -9.967,  -11.891,    0.000,    0.000,    0.000,   -0.090,   -0.090,   -0.112,    0.000,    0.000,    0.000,   -0.293
Os   , S    ,   11.840,   72.317,   72.317,   86.125,    0.000,    0.000,    0.000,    0.655,    0.655,    0.813,    0.000,    0.000,    0.000,    2.123
Os   , BO   ,   11.880,   80.930,   80.930,   51.224,    0.000,    0.000,    0.000,    0.897,    0.897,    0.208,    0.000,    0.000,    0.000,    2.002
Os   , O    ,   12.140,   69.381,   69.381,   82.473,    0.000,    0.000,    0.000,    0.629,    0.629,    0.778,    0.000,    0.000,    0.000,    2.036
Os   , BS   ,   12.960,   93.832,   65.207,   57.745,    0.000,    0.000,    0.000,    1.182,    0.527,    0.320,    0.000,    0.000,    0.000,    2.029
Co   , Vac  ,    1.930,   -4.697,   -4.697,   -3.715,    0.000,    0.000,    0.000,   -0.119,   -0.119,   -0.068,    0.000,    0.000,    0.000,   -0.305
Co   , S    ,    4.310,   24.955,   24.955,   31.679,    0.000,    0.000,    0.000,    0.529,    0.529,    0.836,    0.000,    0.000,    0.000,    1.893
Co   , BO   ,    4.430,   32.672,   32.672,   15.411,    0.000,    0.000,    0.000,    0.913,    0.913,    0.059,    0.000,    0.000,    0.000,    1.884
Co   , BS   ,    4.960,   39.095,   29.016,   14.730,    0.000,    0.000,    0.000,    1.350,    0.576,    0.008,    0.000,    0.000,    0.000,    1.934
Co   , O    ,    5.040,   22.934,   22.934,   39.569,    0.000,    0.000,    0.000,    0.397,    0.397,    1.183,    0.000,    0.000,    0.000,    1.978
Zn   , Vac  ,    0.310,   -3.855,   -3.855,    0.991,    0.000,    0.000,    0.000,   -0.295,   -0.295,    0.498,    0.000,    0.000,    0.000,   -0.092
Zn   , O    ,    1.240,    6.875,    6.875,   13.702,    0.000,    0.000,    0.000,   -0.089,   -0.089,    2.098,    0.000,    0.000,    0.000,    1.919
Zn   , C    ,    1.250,    6.843,    6.195,   13.859,   -0.467,    0.561,   -0.808,   -0.095,   -0.143,    2.156,   -0.069,    0.185,   -0.267,    1.918
Zn   , S    ,    1.370,    6.278,    6.278,   14.999,    0.000,    0.000,    0.000,   -0.180,   -0.180,    2.395,    0.000,    0.000,    0.000,    2.035
Zn   , BO   ,    2.970,   19.986,   19.986,    6.317,    0.000,    0.000,    0.000,    1.073,    1.073,   -0.370,    0.000,    0.000,    0.000,    1.776
Zn   , BS   ,    3.070,   30.173,   12.273,    6.529,    0.000,    0.000,    0.000,    1.810,    0.483,   -0.427,    0.000,    0.000,    0.000,    1.866
Zn_PBEsol, Vac  ,    0.390,   -2.556,   -2.556,   -3.263,    0.000,    0.000,    0.000,   -0.062,   -0.062,   -0.263,    0.000,    0.000,    0.000,   -0.387
Zn_PBEsol, O    ,    1.460,    5.342,    5.342,   24.263,    0.000,    0.000,    0.000,   -0.205,   -0.205,    2.478,    0.000,    0.000,    0.000,    2.067
Zn_PBEsol, S    ,    1.660,    5.109,    5.109,   25.349,    0.000,    0.000,    0.000,   -0.237,   -0.237,    2.607,    0.000,    0.000,    0.000,    2.132
Zn_PBEsol, BO   ,    3.310,   17.911,   17.911,   15.845,    0.000,    0.000,    0.000,    0.569,    0.569,    1.064,    0.000,    0.000,    0.000,    2.203
Cd   , Vac  ,    0.100,   -4.338,   -4.338,   -0.346,    0.000,    0.000,    0.000,   -0.403,   -0.403,    0.531,    0.000,    0.000,    0.000,   -0.274
Cd   , C    ,    0.930,    9.658,    8.525,   11.521,    0.347,    0.981,    0.600,    0.184,   -0.051,    1.792,    0.144,    1.185,    0.725,    1.926
Cd   , O    ,    0.960,    9.247,    9.247,   11.648,    0.000,    0.000,    0.000,    0.073,    0.073,    1.804,    0.000,    0.000,    0.000,    1.950
Cd   , S    ,    1.030,    9.054,    9.054,   12.606,    0.000,    0.000,    0.000,   -0.015,   -0.015,    2.090,    0.000,    0.000,    0.000,    2.060
Cd   , BO   ,    2.150,   21.451,   21.451,    6.373,    0.000,    0.000,    0.000,    1.655,    1.655,   -1.372,    0.000,    0.000,    0.000,    1.939
Cd_PBEsol, Vac  ,    0.210,   -4.719,   -4.719,    0.706,    0.000,    0.000,    0.000,   -0.308,   -0.308,    0.428,    0.000,    0.000,    0.000,   -0.188
Cd_PBEsol, O    ,    1.310,    9.835,    9.835,   16.713,    0.000,    0.000,    0.000,    0.022,    0.022,    1.849,    0.000,    0.000,    0.000,    1.894
Cd_PBEsol, S    ,    1.470,    9.479,    9.479,   17.856,    0.000,    0.000,    0.000,   -0.038,   -0.038,    2.046,    0.000,    0.000,    0.000,    1.970
Cd_PBEsol, C    ,    2.750,   15.503,   16.578,   30.941,   -0.227,   -0.931,   -0.393,   -0.137,   -0.041,    3.571,   -0.040,   -0.427,   -0.180,    3.393
Cd_PBEsol, BO   ,    2.760,   24.471,   24.471,    7.861,    0.000,    0.000,    0.000,    1.203,    1.203,   -0.481,    0.000,    0.000,    0.000,    1.925
Cd_PBEsol, BS   ,    2.800,   36.101,   17.001,    8.341,    0.000,    0.000,    0.000,    2.161,    0.463,   -0.550,    0.000,    0.000,    0.000,    2.073
Tl   , Vac  ,    0.440,   -2.281,   -2.281,   -2.188,    0.000,    0.000,    0.000,   -0.179,   -0.179,   -0.083,    0.000,    0.000,    0.000,   -0.441
Tl   , S    ,    0.920,    7.742,    7.742,   12.400,    0.000,    0.000,    0.000,    0.375,    0.375,    0.953,    0.000,    0.000,    0.000,    1.703
Tl   , O    ,    0.950,    7.565,    7.565,   12.663,    0.000,    0.000,    0.000,    0.341,    0.341,    1.005,    0.000,    0.000,    0.000,    1.688
Tl   , BO   ,    0.980,    7.845,    7.845,   10.320,    0.000,    0.000,    0.000,    0.485,    0.485,    0.663,    0.000,    0.000,    0.000,    1.632
Tl   , BS   ,    1.040,   11.760,    6.331,    9.280,    0.000,    0.000,    0.000,    2.279,   -0.917,    0.410,    0.000,    0.000,    0.000,    1.773
\end{filecontents*}

\preprint{APS/123-QED}

\title{Review of point defect structures in hexagonal close packed metals and across the  Periodic Table}%

\author{Andrew R. Warwick}
\affiliation{%
 UK Atomic Energy Authority, Culham Science Centre, Oxfordshire, OX14 3DB, United Kingdom
}%
\author{Pui-Wai Ma}
\author{Sergei L. Dudarev}
\affiliation{%
 UK Atomic Energy Authority, Culham Science Centre, Oxfordshire, OX14 3DB, United Kingdom
}%

\date{May 26, 2025}

\begin{abstract}
We present a comprehensive \textit{ab initio }dataset of formation energies and elastic properties of intrinsic point defects across all the transition and rare earth hexagonal close packed (hcp) metals, as well as metalloid elements with hcp crystal structure. Point defect properties appear weakly correlated with the $c/a$ ratio of the hcp lattice. Instead, it is the position of an element in the Periodic Table that primarily defines the relaxation volume tensor, elastic dipole tensor and formation energy of a point defect. This suggests that the local variations in the electronic structure and interatomic bonding at the core of a defect dominate its properties, as opposed to long-range elastic deformations. Across all the metals, we find that the relaxation volumes of vacancies and self-interstitial defects are correlated with atomic volumes, with values of -0.35 and 1.46 atomic volumes for a vacancy and a self-interstitial defect, respectively, providing a universally good approximation independent of the crystal structure.  This study complements and completes the existing point defect databases spanning body-centred cubic and face-centred cubic metals.
\end{abstract}

\maketitle

\section{Introduction}

Properties of metals are sensitive to perturbations of their underlying crystal structure. %
Intrinsic point defects are the simplest types of lattice defects, involving the removal or insertion of a single atom. %
Vacancies and self-interstitial atom (SIA) defects are readily created, for example, by high-energy particle irradiation in a nuclear reactor. %
Their coalescence into larger, extended defects, such as dislocations and voids, is among the causes of hardening and embrittlement. %
The accumulation of point defects also gives rise to dimensional changes and swelling. 

Defects interact through long-range elastic fields, with the strength of interaction determined by elastic dipole tensors $P_{ij}$ \cite[p. 161]{Leibfried1978}. %
A sample of the host metal, free from body forces and tractions, changes shape and volume according to the relaxation volume tensor ${\Omega_{ij} = S_{ijkl} P_{kl}}$ \cite{Ma2019b,Ma2021} of the defects contained in it, where $S_{ijkl}$ is the elastic compliance tensor \cite[p. 29]{Sutton2020}. %
Models for microstructural evolution at larger length scales, treating interactions between the defects, require these tensorial quantities as input. %
Compiling a database of point defect elastic tensors is a necessary step involved in predicting deformations of materials operating in an irradiation environment. %

\begin{figure}
    \includegraphics[width=0.48\textwidth]{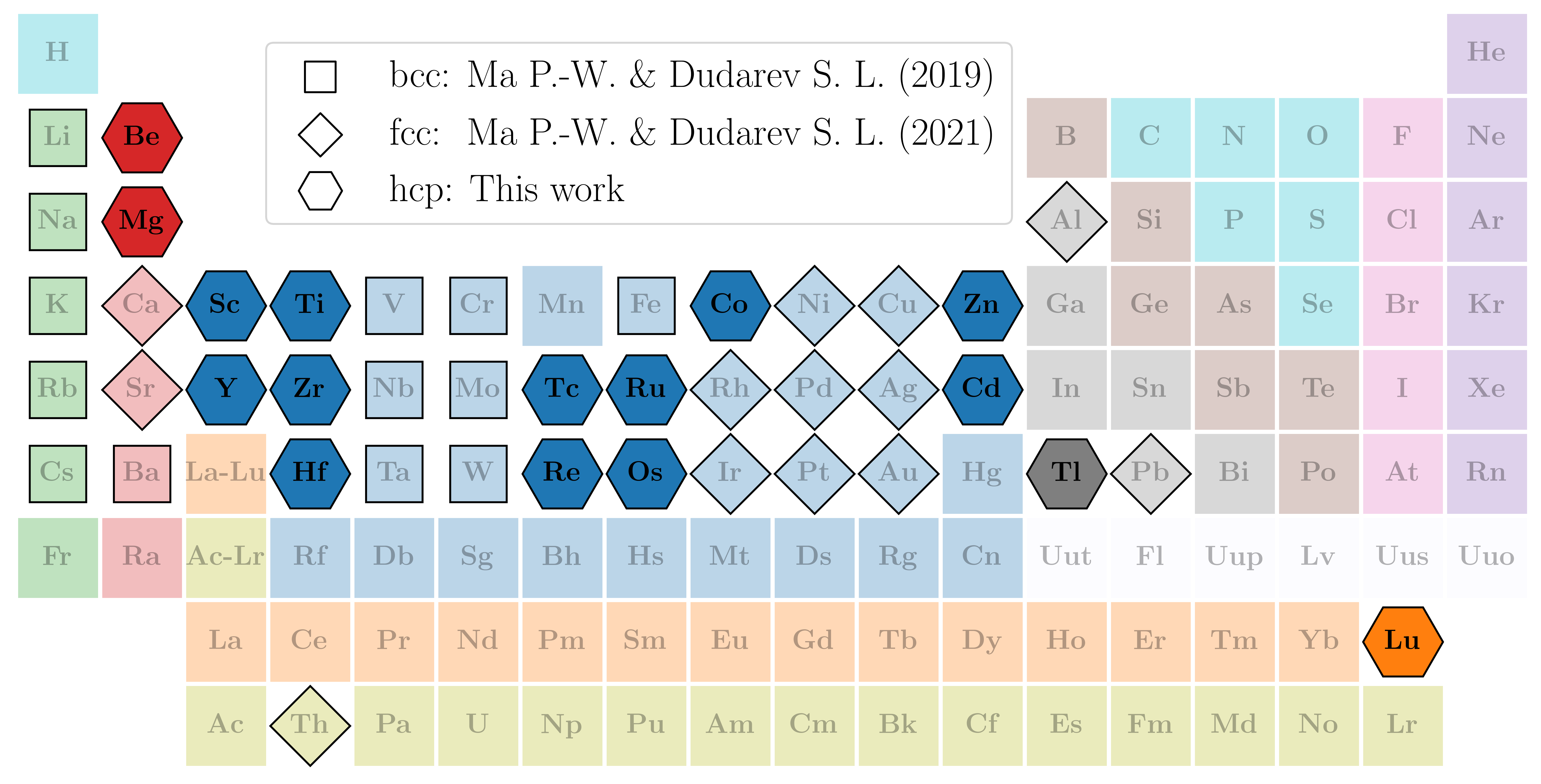}
    \caption{The elements with hcp structure considered in this work and highlighted by hexagon cells. The bcc \cite{Ma2019b} and fcc \cite{Ma2021} metals considered in previous studies are depicted by the bordered squares and diamonds, respectively. Most of the lanthanide and actinide hcp metals were not included in this study.}
    \label{§1-fig:periodic_table}
\end{figure}

$P_{ij}$ and $\Omega_{ij}$ for a vacancy and various SIA configurations have been computed \textit{ab initio} and tabulated across all the body centred cubic (bcc) \cite{Ma2019b} and face centred cubic (fcc) \cite{Ma2021} metals, save the lanthanides and actinides. %
The lowest energy SIA structure in all the bcc metals has the $\langle 1 1 1 \rangle$ symmetry, bar magnetic Fe \cite{NguyenManh2006}. In Cr, Mo and W the $\langle 111 \rangle $ symmetry is slightly broken by the buckling occurring in the core of the defect \cite{Olsson2009,Ma2019}. In fcc metals, the $\langle 100 \rangle$ dumbbell is often, but not always, the lowest energy SIA configuration. %

A compilation of point-defect elastic dipole and relaxation volume tensors for the hexagonal close-packed (hcp) metals is currently missing. %
Many hcp metals play an important role in engineering applications such as lightweight structural components composed of Mg \cite{Prasad2022} or Ti \cite{Williams2020}, Zr nuclear fuel claddings \cite{Adamson2019}, and Be as a candidate neutron multiplier material in nuclear fusion reactors \cite{Wang2025}. %

Until now, \citet{Pasianot2016} has conducted the most comprehensive \textit{ab initio} study of elastic properties of point defects in hcp metals. %
The trends in the defect formation energy, dipole tensors, and relaxation volumes with respect to the $c/a$ ratio were explored and compared to resistivity recovery, internal friction, and Huang diffuse X-ray scattering (DXS) data. %
Seven elements were included in the study \cite{Pasianot2016}: the two elements that have a larger than ideal $c/a$ ratio, Cd and Zn, as well as Be, Ti, Zr, Mg and Co. %
For Be, Ti and Zr, the lowest formation energy SIA was found to be the basal octahedral (BO) configuration, with Mg and Co instead favouring the crowdion (C) and $[0001]$ split dumbbell (S) defect structure, respectively. %
The SIA formation energies computed for Cd and Zn showed a clear preference for the out-of-basal-plane configurations over the basal SIAs. %
Due to the ease of transformation of the ground state configuration of the defect into alternative configurations, it was suggested that SIA migration is likely to be three-dimensional in Zn, Cd and Mg. %
The predicted values of $P_{ij}$ were compared with Huang diffuse X-ray scattering (DXS) measurements, suggesting that the computed and experimentally measured values were often not compatible with each other. %
Dipole tensor calculations have also been performed for the group IV hcp metals Ti, Zr and Hf \cite{Verite2013}, showing that in Zr the SIA configurations with monoclinic symmetry had the lowest energy and were consistent with internal friction experimental measurements. %
The elastic properties of defects across the Periodic Table have also been investigated using EAM potentials \cite{Oh1989,Monti1991,Hu2003}, which often exhibit relatively poor accuracy in comparison with {\it ab initio} calculations. %

Most studies of defects in hcp metals focussed on formation and migration energies \cite{Angsten2014}. %
Of particular note are the \textit{ab initio} high throughput calculations conducted by \citet{Shang2016} that have yielded a complete database of energies for vacancy defects across \textit{all} the elements in the Periodic Table, including the elements with hcp crystal structure. %
Self-interstitial atom defects have been investigated, using {\it ab initio} methods, in Zn \cite{Nitol2021}, Be \cite{Ganchenkova2009, Ferry2019}, Y \cite{Borodin2019a}, Ti \cite{Wooding1995,Wen2000,TundeRaji2009,Verite2013,Nayak2018}, Zr \cite{Ong1982, DeDiego2002, Domain2005, DeDiego2008, DeDiego2011, Samolyuk2013, Peng2012, Verite2013, Peng2013, Christensen2015, Samolyuk2014, Peng2015, Varvenne2017}, Mg \cite{Fendrik1982, Li2015}, Dy \cite{Fan2015}, Hf \cite{Verite2013} and Sc \cite{Fan2011}. %
The results are broadly consistent with an earlier hypothesis that the SIAs lying in the basal plane are the most stable configurations in all the hcp elements with the lower than ideal $c/a$ ratio \cite{Bacon1993}. %

In this study, we compute the formation energies, dipole tensors and relaxation volume tensors of all the hcp metals and hcp metalloid elements in the Periodic Table, with the exception of lanthanides and actinides, shown in Figure \ref{§1-fig:periodic_table} by the hexagon symbols. %
The bcc \cite{Ma2019b,Ma2019} and fcc \cite{Ma2021} metals studied earlier are denoted by squares and diamonds, respectively. %
We outline the point defect configurations and simulation methods in section \ref{sec:methodology}, followed by the tabulation and analysis of our results in section \ref{sec:results}, with conclusions provided in section \ref{sec:conclusions}. %

\section{Methodology}
\label{sec:methodology}

\begin{table}
\centering
\begin{ruledtabular}
\begin{tabular}{cccc}
Element &
Valence electrons &
$E_{cut}$ (eV) & 
$k$-grid %
\\ \hline \\
Be & $1s^2 2s^2$     & 800  &  $19\times19\times12$ \\
Mg & $2s^2 2p^6 3s^2$     & 1000 &  $17\times17\times11$ \\
Sc & $3s^2 3p^6 4s^1 3d^2$       & 700  &  $11\times11\times07$ \\
Y  & $4s^2 4p^6 5s^1 4d^2$     & 400  &  $17\times17\times11$ \\
Lu & $5s^2 5p^6 4f^{14} 6s^2 5d^1 $   & 600  &  $13\times13\times08$ \\
Ti & $3s^2 3p^6 4s^1 3d^3$       & 600  &  $15\times15\times09$ \\
Zr & $4s^2 4p^6 5s^1 4d^3$     & 700  &  $17\times17\times11$ \\
Hf & $5s^2 5p^6 5d^4$    & 700  &  $17\times17\times11$ \\
Tc & $4p^6 4d^6 5s^1$       & 800  &  $19\times19\times12$ \\
Re & $5p^6 6s^1 5d^6$      & 600  &  $19\times19\times12$ \\
Ru & $4s^2 4p^6 4d^7 5s^1$       & 900  &  $17\times17\times11$ \\
Os & $5p^6 6s^1 5d^7 $       & 600  &  $17\times17\times11$ \\
Co & $3s^2 3p^6 4s^1 3d^8$     & 900  &  $19\times19\times12$ \\
Zn & $3d^{10} 4s^2$  & 700  &  $17\times17\times11$ \\
Cd & $4d^{10} 5s^2$  & 700  &  $17\times17\times09$ \\
Tl & $5d^{10} 6s^2 6p^1$       & 900  &  $17\times17\times11$ \\
\end{tabular}
\end{ruledtabular}
\caption{Parameters for atomic relaxation simulations  performed using VASP in the PBE approximation for the exchange-correlation functional. $E_{cut}$ is the plane wave cut-off energy and $k$-grid denotes the $\Gamma$-centered evenly spaced $k$ point mesh utilized for a two-atom hcp primitive cell. }
\label{table:VASP_parameters}
\end{table}%

\begin{figure}
    \centering
    \includegraphics[width=0.45\textwidth]{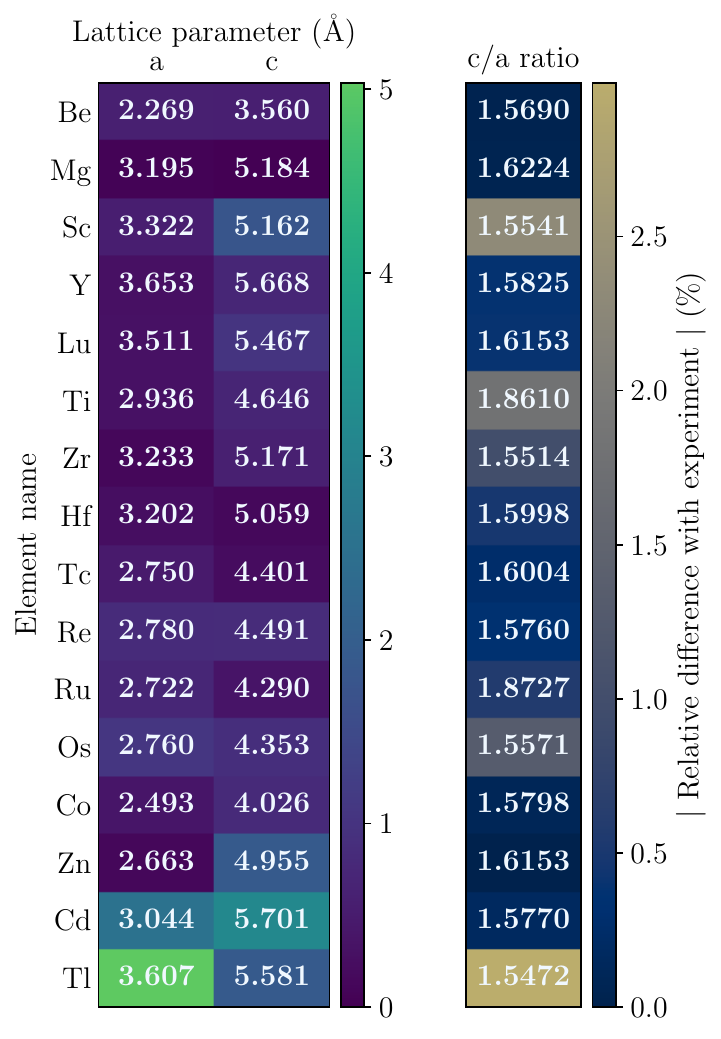}
    \caption{\textit{Ab initio} predicted values of lattice parameters and $c/a$ ratios for all the hcp elements included in this study. The cells in the table are coloured according to the absolute difference between the calculated values and experimental measurements. Experiment data are sourced from Ref. \cite{Arblaster2018} and are the 0K values, with the exception of Hf for which the lowest temperature experimental values correspond to room temperature.}
    \label{§2-fig:lattice_param_and_ca_comparison}
\end{figure}

Density functional theory (DFT) calculations were performed using the \textit{Vienna Ab-Initio Simulation Package} (\textsc{vasp}) version 6.4.1 \cite{Kresse1993,Kresse1994a,Kresse1996a,Kresse1996b}. %
In Table \ref{table:VASP_parameters} we summarise the selected plane wave energy cut-off $E_{cut}$, \textit{k}-point mesh and valence electron configuration of the plane augmented wave (PAW) pseudopotential \cite{Kresse1994b,Kresse1999} for each hcp metal. %
For a given element, we have chosen the pseudopotential with the lowest number of core electrons, to achieve the highest possible accuracy in the treatment of interatomic interactions at short distances in the SIA structures. %
The results of convergence tests are given in the Supplementary Material. %
The energies computed by the self consistent field method were converged to within \unit{10^{-8}}{eV} and the point defect structures were relaxed until the maximum magnitude of force on any atom was smaller than \unit{0.01}{eV/\AA}. %
The Perdew-Burke-Ernzerhof (PBE) exchange-correlation functional $E_{xc}$ \cite{Perdew1996} was employed for all the hcp elements. %
For Be, Zn and Cd, the PBEsol functional \cite{Perdew2008} was also considered, as discussed in section \ref{sec:results}. %
Otherwise, earlier studies have shown that various choices for $E_{xc}$ often do not have a significant effect on the resulting elastic dipole tensors \cite{Ma2019b,Ma2021}. %

Bulk lattice parameters were optimised such that the change in energy between the successive ionic relaxation steps in calculations of perfect lattice structures was less than \unit{10^{-7}}{eV}. %
As shown in Figure \ref{§2-fig:lattice_param_and_ca_comparison}, the predicted  lattice parameters are in good agreement with experiment data \cite{Arblaster2018}, and the relative error of either $a$ or $c$ with respect to experiment does not exceed 5\%. %
Furthermore, the predicted $c/a$ ratios differ from experimental values by less than 3\%. %

\begin{figure}
    \centering
    \includegraphics[width=0.45\textwidth]{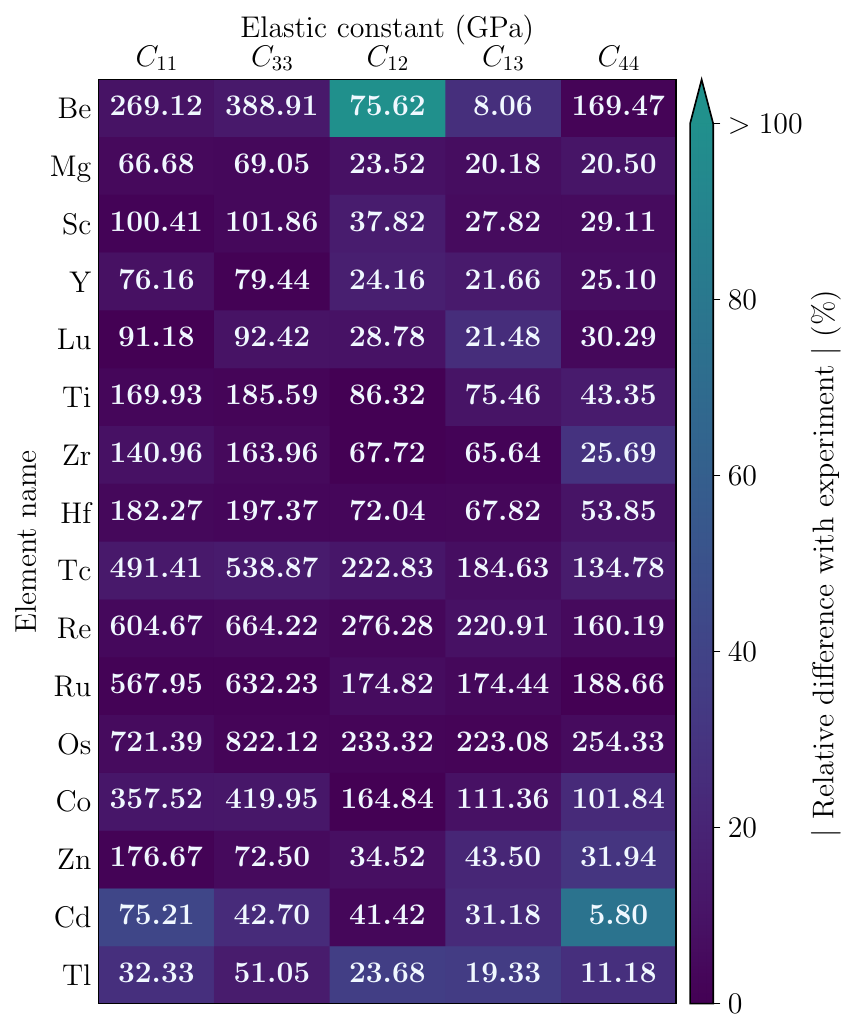}
    \caption{\textit{Ab initio} predicted values of elastic constants for all the hcp elements included in this study. Cells in the table are coloured according to the absolute difference between the calculated values and experimental measurements. Experiment data are sourced from \citet{Simmons1971} for temperatures less than 4K for all the elements, with the exception of Lu \cite{Tonnies1971} (4K), Os \cite{Pantea2009} (5K), Sc \cite{Leisure1993} (300 K) and Tc \cite{Guillermet1989} (298 K).}
    \label{§2-fig:elastic_constants_comparison}
\end{figure}

\begin{figure*}
    \centering
    \includegraphics[width=1.\textwidth]{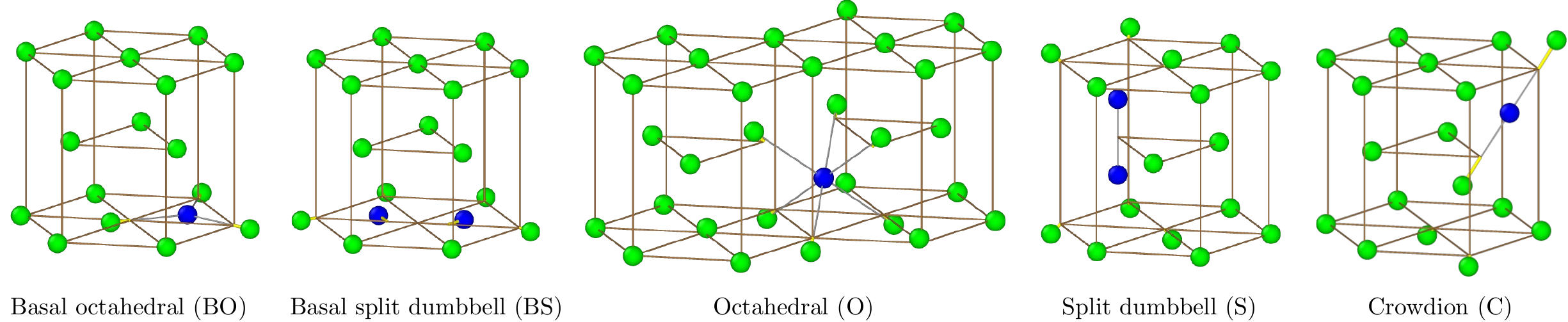}
    \caption{From left to right: relaxed configurations of the basal octahedral (BO), basal $[2\bar1\bar10]$ split dumbbell (BS), octahedral (O), $[0001]$ split dumbbell (BS) and crowdion (C) SIAs for Zr. The inserted atom, or a pair of dumbbell atoms sharing an atomic site, are highlighted in blue. Atomic displacements from the initial configuration are indicated by yellow arrows.}
    \label{§2-fig:SIA_configurations}
\end{figure*}

The elastic constants were computed \textit{via} the symmetrised finite differences method \cite{Le_Page2002, Wu2005}. %
In general, the relative differences between the predicted and measured elastic constants \cite{Simmons1971, Tonnies1971, Pantea2009, Leisure1993, Guillermet1989} are satisfactory. %
The largest relative error appears in the $C_{12}$ elastic constant of Be, for which the experimental value is \unit{C_{12} = 27.6}{GPa} \cite{Simmons1971}. %
We note, however, that the absolute error 
in this case is still similar to the scale of discrepancies found across all the hcp elements. %

Ferromagnetic hcp Co was treated in the collinear magnetic approximation, similarly to the treatment of ferromagnetic bcc Fe and antiferromagnetic bcc Cr \cite{Ma2019b}, and ferromagnetic fcc Ni \cite{Ma2021}. %
The calculated magnetic moment of 1.616 $\mu_B$ per Co atom is in good agreement with the experimentally measured value of 1.71 $\mu_B$ per atom \cite{Myers1997}. %

The formation energies $E^{\mathrm{F}}_{\mathrm{def}}$ and elastic dipole tensors of a vacancy and five self interstitial atom defect configurations were computed for each chemical element. %
The formation energy of a defect is defined as
\begin{equation}
    E^{\mathrm{F}}_{\mathrm{def}} = \left[ E_{\mathrm{def}}(N_{\mathrm{def}}) - E^{\mathrm{app}} \right] - \frac{N_{\mathrm{def}} }{N_{\mathrm{perf}}}E_{\mathrm{perf}}(N_{\mathrm{perf}}) - E^{\mathrm{corr}}_{\mathrm{el}},
\end{equation}
where the energy of a simulation cell containing a defect in a configuration of the total of $N_{\mathrm{def}}$ atoms is $E_{\mathrm{def}}$, and the corresponding perfect lattice reference cell containing $N_{\mathrm{perf}}$ atoms is $E_{\mathrm{perf}}$. %
For a vacancy or a self interstitial atom defect, ${N_{\mathrm{def}} - N_{\mathrm{perf}}}$ is $-1$ or $+1$ respectively. %
$E^{\mathrm{app}}$ is the energy associated with  externally applied strain and $E^{\mathrm{corr}}_{\mathrm{el}}$ is the energy of elastic interaction between a defect and all the periodic images of it. %
Accounting for the energy of interaction with periodic images of a defect structures $E^{\mathrm{corr}}_{\mathrm{el}}$ is possible \cite{Dudarev2018PRM}, and the corresponding correction term was computed using the \textsc{calanie} program \cite{Ma2020}. %
Further details may be found in Refs. \cite{Ma2019b,Ma2021}. %

We constructed simulation cells from repeats of the 2 atom hcp primitive cell with space group 194 $P6_3/mmc$ and occupied Wyckoff position ${c = (1/3, 2/3, 1/4)}$ \cite[ch. 2.3]{Aroyo2016}. %
Both atoms in the primitive cell belong to the orbit of a single Wyckoff position. %
It follows that the structure and tensorial properties of a vacancy at either site in the primitive cell are related by a symmetry operation of the crystal space group. %
Thus, there is only one type of vacancy defect in an hcp crystal. %
Conversely, multiple non-equivalent SIA geometries need to be considered. %
The relaxed configurations of the basal octahedral (BO), basal $\langle 2\bar1\bar10 \rangle$ split dumbbell (BS), octahedral (O), $\langle 0001 \rangle$ split dumbbell (BS) and crowdion (C) SIAs for Zr are shown in figure \ref{§2-fig:SIA_configurations}. %
\citet{Verite2013} and \citet{Pasianot2016} have highlighted the important role of low-symmetry defect configurations in determining the migration pathways of SIAs. %
For certain low symmetry configurations, the formation energy was found to be only slightly above the lowest energy high symmetry SIA configuration. %
Here, we present only the high symmetry initial configurations so that the configurational and energy landscape of SIAs across all the hcp metals may be explored efficiently. %
As reported by \citet{Pasianot2016}, we often found that the high symmetry tetrahedral, basal tetrahedral and basal crowdion defects are unstable or have relatively high formation energies. %
These configurations were not included in this study. %

\begin{figure}
    \centering
    \includegraphics[width=0.45\textwidth]{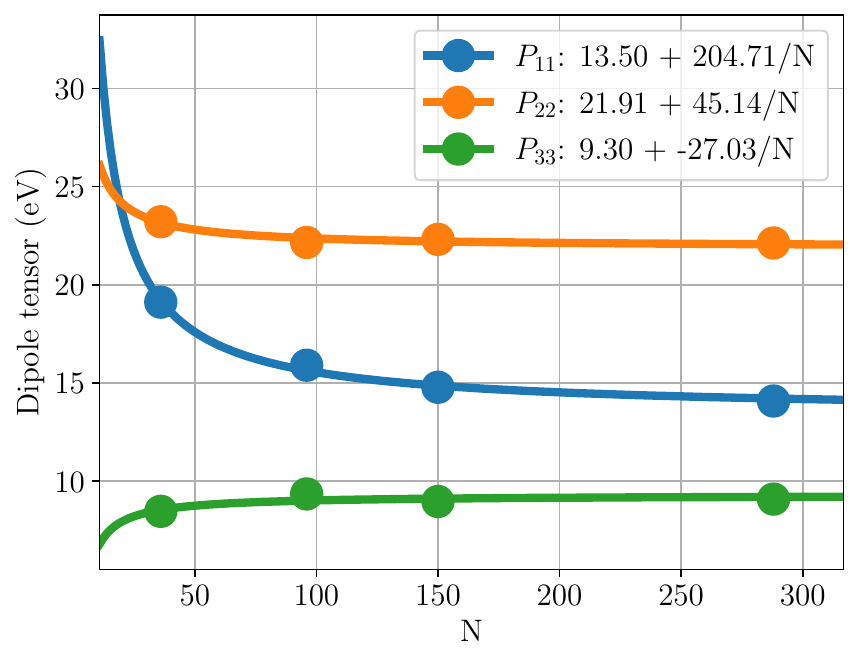}
    \caption{ Example linear extrapolation of dipole tensor components with respect to $N$, where $N$ denotes the number of atoms in the reference supercell, for a basal ${[2\bar1 \bar1 0]}$ split dumbell SIA defect in Zr. Cell shape and volume are fixed during internal atomic relaxation. The $1/N$ variation comes from the fact that interaction of a defect with its periodically translated images varies as the inverse cube of linear dimensions of the cell \cite{Varvenne2013,Dudarev2018PRM}. }
    \label{§2-fig:1/N_extrapolation}
\end{figure}

Consider a simulation cell containing a single point defect in a cell containing the total of $N_{def}$ atoms. %
The structure of the defect is relaxed under the condition that the cell shape and size are fixed. %
The dipole tensor ${P_{ij}}(N_{def})$ is then evaluated from the average residual stress in the cell $\bar{\sigma}_{ij}$ as \cite{Clouet2008,Dudarev2018PRM}
\begin{align}
    P_{ij}(N_{def}) = - V\bar{\sigma}_{ij},
\end{align}
where $V$ is the volume of the cell. 
Asymptotically, the stress field of a defect varies as the inverse cube of the distance from its centre $||\sigma _{ij}({\bf r})|| \sim 1/r^3$ \cite{Sutton2020}. %
Hence, each defect configuration was relaxed using at least three cells containing different number of atoms $N$. %
The dipole tensor components were then extrapolated to the limit of infinite cell size to obtain the asymptotic values $P^\infty_{ij}$ \textit{via} a least squares fit with respect to a linear function of $1/N$ such that \cite{Varvenne2013,Varvenne2017}
\begin{align}
    P_{ij}(N) = \frac{\alpha_{ij}}{N} + P^\infty_{ij},
\label{eq:dipole_1N}
\end{align}
where $\alpha_{ij}$ is a fitted constant. %
Often we found that the reference supercell representing the   relaxed perfect crystal contains some remnant stress, stemming from the approximations involved in the numerical relaxation procedure. %
In this way, we have computed the stress of the initial perfect crystal supercell and subtracted it from the elastic interaction-corrected dipole tensor for each cell size before performing the $1/N$ extrapolation. %
The number of repeats along each lattice vector were chosen such that the cell edge lengths were approximately equal, resulting in 36, 96 and 150 atom cells arising from the $3 \times 3 \times 2 $, $4 \times 4 \times 3 $, $5 \times 5 \times 3 $ unit cell repeats respectively. %
Where computational resources were available, relaxations were also performed on larger cell sizes. %
Due to the large $c/a$ ratio of Zn and Cd, a 64-atom $4 \times 4 \times 2 $ cell was considered in place of a 96 atom $4 \times 4 \times 3 $ cell. %
For many of the transition hcp metals, we found that Eq.\ \ref{eq:dipole_1N} provided a good fit to the point defect data. %
However, Be, Mg, Zn, Cd and Tl presented less optimal fits. %
In order to maintain consistency, we have presented the result of applying the procedure described above to all elements. %

Relaxation volume tensors of isolated point defects $\Omega_{ij}$ were computed from the corresponding extrapolated dipole tensors $P^\infty_{ij}$ \textit{via}
\begin{align}
    \Omega_{ij} = S_{ijkl} P^{\infty}_{ij}.
\end{align}
Unless stated otherwise, we express $\Omega_{ij}$ in the atomic volume units $\Omega_0$ for the respective hcp metal, where for the hcp crystal structure ${\Omega_0 = \frac{\sqrt{3}}{4} a^2 c}$. %
Example extrapolations for the elastic dipole tensors of a Zr vacancy and a Zn crowdion are shown in figure \ref{§2-fig:1/N_extrapolation}. %
Extrapolations for other elements and defect configurations are given in the Supplementary Material. %

\section{Results}
\label{sec:results}

\begin{figure}
    \centering
    \includegraphics[width=0.48\textwidth]{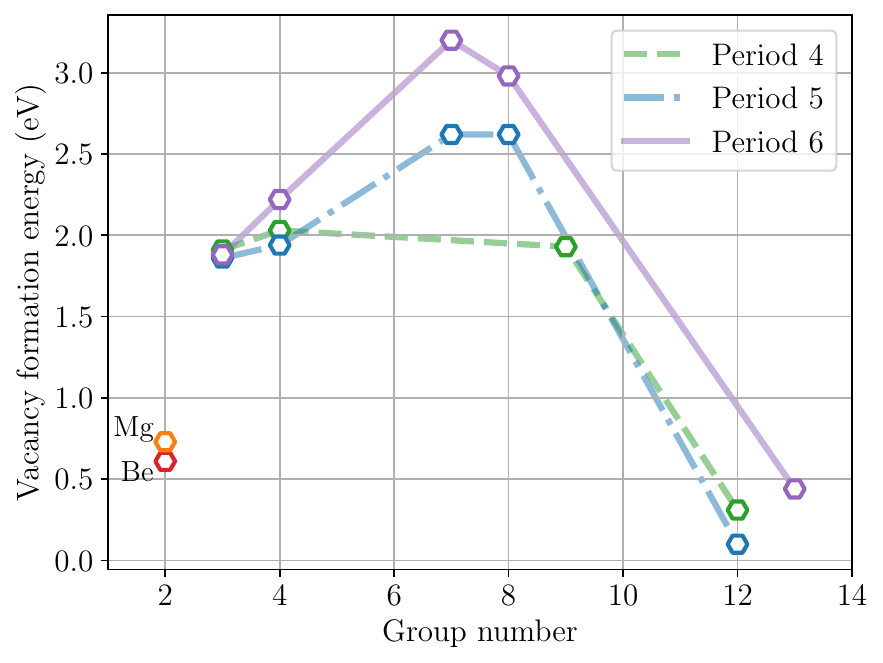}
    \caption{Formation energies of vacancies in hcp metals plotted as a function of the group number of the respective chemical element in the Periodic Table.}
    \label{§3-fig:vacancy_Ef}
\end{figure}

The formation energies $E_f$, elastic dipole tensors $P_{ij}$ and relaxation volume tensors $\Omega_{ij}$ for point defects across all hcp metals highlighted in FIG. \ref{§1-fig:periodic_table} are summarised in Tables II - XX. %
In sections \ref{sec:results-subsec:vacancies} and \ref{sec:results-subsec:SIA}, we analyse the properties of vacancies and SIAs respectively. %
Our values for the dipole and relaxation volume tensors of defects included in the study by \citet{Pasianot2016} agree with the values computed there. %

\subsection{Vacancies}
\label{sec:results-subsec:vacancies}

\begin{figure}
    \centering
    \includegraphics[width=0.45\textwidth]{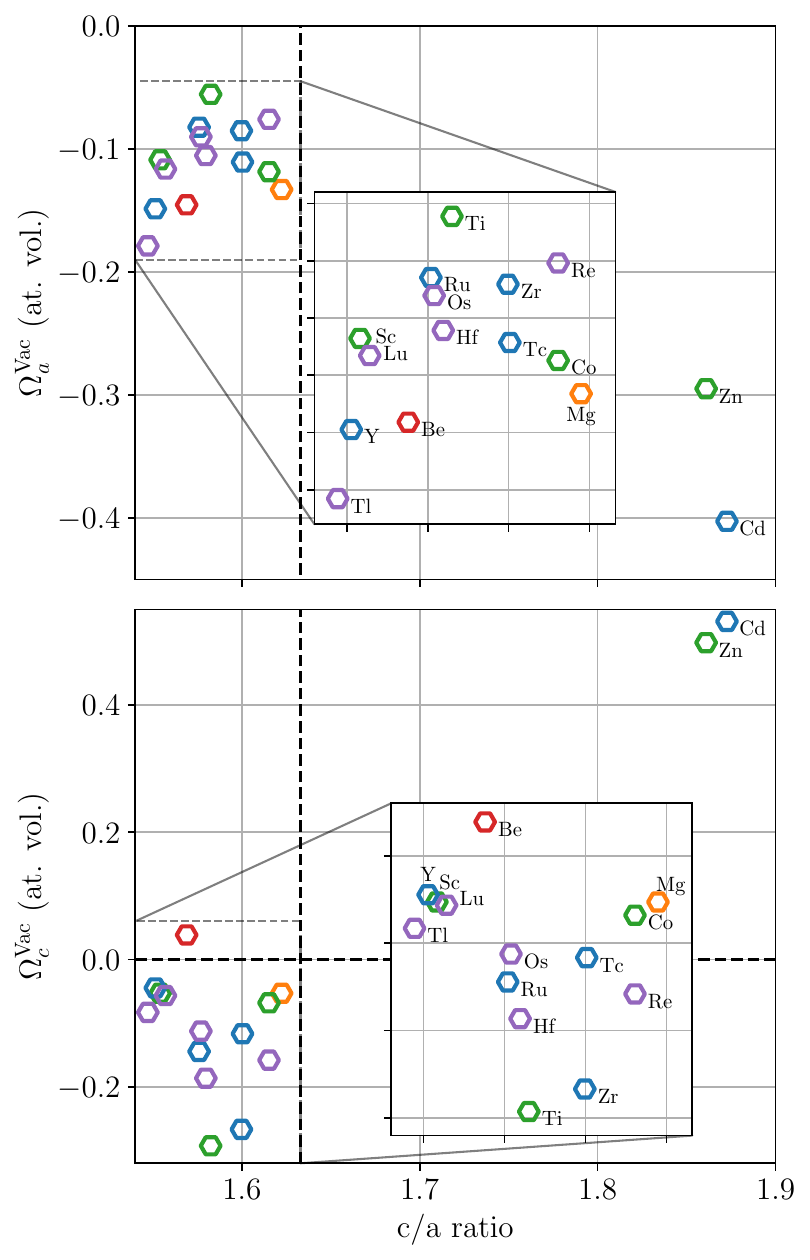}
    \caption{Vacancy relaxation volume tensor components $\Omega_a$ and $\Omega_c$, defined in equation \ref{eq:vacancy_Oij} with respect to the hcp crystallographic coordinate system. Vertical dashed lines indicate the ideal ratio $c/a = \sqrt{8/3}$.}
    \label{§3-fig:vacancy_rel_vol}
\end{figure}

\begin{figure}
    \centering
    \includegraphics[width=0.45\textwidth]{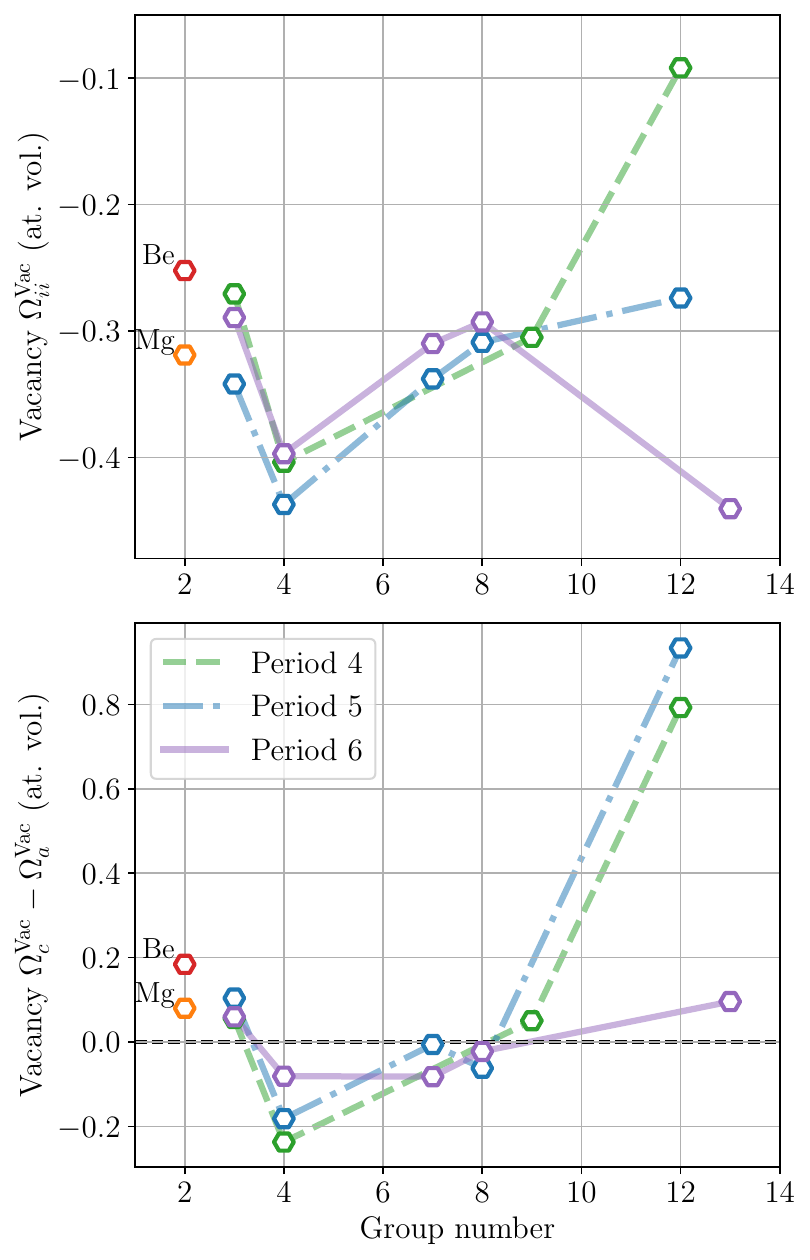}
    \caption{Vacancy relaxation volume tensor components $\Omega_a$ and $\Omega_c$, defined in equation \ref{eq:vacancy_Oij} with respect to the hcp crystallographic coordinate system. Vertical dashed lines indicate the ideal ratio $c/a = \sqrt{8/3}$.}
    \label{§3-fig:vacancy_Oij_relvol_and_deviator}
\end{figure}

The vacancy formation energies found in our study agree with those computed by \citet{Shang2016} within 0.2 eV for all the elements except beryllium where our value for the vacancy formation energy \unit{E^{\mathrm{Vac}}_F = 0.16}{eV} is significantly lower than \unit{0.92}{eV} found in Ref. \cite[Table 3]{Shang2016}. %
The discrepancy may be a result of our choice of $E_{cut}$, \textit{k}-point mesh density or choice of Be pseudopotential. %
Here we used ${E_{cut} = 800 \, \mathrm{eV}}$ compared to ${E_{cut} = 1.4\times \mathrm{ENMAX} = 345 \, \mathrm{eV}}$ where ENMAX is the recommended cut-off supplied in the Be POTCAR file, or 432 eV for the Be\_sv POTCAR. %

The relaxation volume tensor of a vacancy in an hcp crystal has the form \cite{Varvenne2017}
\begin{align}
    \boldsymbol{\Omega}^{\mathrm{Vac}} = 
    \begin{pmatrix} 
        \Omega^{\mathrm{Vac}}_{a} & 0 & 0 \\
        0 & \Omega^{\mathrm{Vac}}_{a} & 0 \\
        0 & 0 & \Omega^{\mathrm{Vac}}_{c} \\
    \end{pmatrix},
    \label{eq:vacancy_Oij}
\end{align}
for independent strains in the basal plane $\Omega^{\mathrm{Vac}}_a$ and along the $c$-axis $\Omega^{\mathrm{Vac}}_c$. %
$\Omega^{\mathrm{Vac}}_a$ and $\Omega^{\mathrm{Vac}}_c$ are plotted in figure \ref{§3-fig:vacancy_rel_vol} as a function of $c/a$ ratio. %
For elements with the less than ideal $c/a$ ratio, we observe a weak positive correlation. %
We also note that $\Omega^{\mathrm{Vac}}_c$ is found to be positive for Cd and Zn, despite the fact that vacancies in materials are centres of contraction.
For Be, the magnitude of ${\Omega^{\mathrm{Vac}}_{c} = 0.039 \Omega_0}$ is smaller than the errors associated with the choice of the exchange-correlation functional $E_{xc}$ and the pseudopotential. Hence, the sign of the vacancy $c$ strain in beryllium is uncertain.

In general, a second rank tensor $\boldsymbol{\Omega}$ may be decomposed into an isotropic ${\frac{1}{3}\mathrm{Tr}(\boldsymbol{\Omega})\mathbf{I}}$ and deviatoric ${\boldsymbol{\Omega}^d = \boldsymbol{\Omega} - \frac{1}{3}\mathrm{Tr}(\boldsymbol{\Omega})\mathbf{I}}$ parts, where $\mathbf{I}$ is the Kronecker delta tensor. %
Hence, we may rewrite eq.\ \ref{eq:vacancy_Oij} as
\begin{align}
    \boldsymbol{\Omega}^{\mathrm{Vac}} = 
    \frac{\Omega^{\mathrm{Vac}}_{ii}}{3} \mathbf{I}
    +
    \frac{\Omega^{\mathrm{Vac}}_{c}-\Omega^{\mathrm{Vac}}_{a}}{3} \begin{pmatrix} 
        -1 & 0 & 0 \\
        0 & -1 & 0 \\
        0 & 0 & 2 \\
    \end{pmatrix},
\end{align}
where the relaxation volume $\Omega^{\mathrm{Vac}}_{ii}$ is the trace of tensor $\boldsymbol{\Omega}^{\mathrm{Vac}}$, and the principal shear strain ${\Omega^{\mathrm{Vac}}_{c}-\Omega^{\mathrm{Vac}}_{a}}$ is a measure of the difference in the $c$ and $a$ strains induced by a vacancy, and is proportional to the magnitude of the deviatoric part of $\boldsymbol{\Omega}^{\mathrm{Vac}}$. %
The volumetric and shear strains are plotted in figure \ref{§3-fig:vacancy_Oij_relvol_and_deviator} with respect to the group number of the element in the Periodic Table. %
The dilatation and shear strains become more negative in elements of group IV, increasing again across a given period. %
Once again, the cases of Cd and Zn are anomalous, with the ${\Omega_c - \Omega_a}$ value being relatively large due to the substantial positive $c$ strain. %

\begin{figure}
    \centering
    \includegraphics[width=0.48\textwidth]{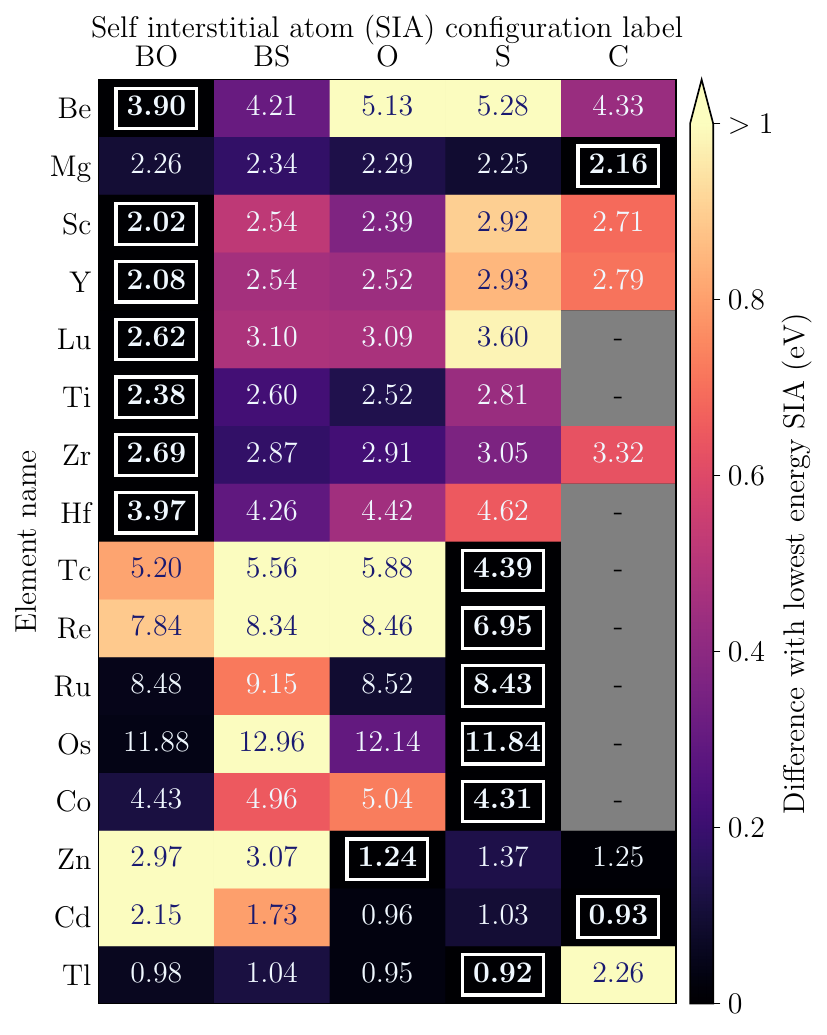}
    \caption{Table of SIA formation energies across the hcp elements in the Periodic Table. For a given element, cells are coloured according to the difference between the formation energy of a configuration and the energy of the most stable SIA structure, highlighted in boldface and white borders. }
    \label{§3-fig:SIA_formation_energy}
\end{figure}

\subsection{Self interstitial atoms}
\label{sec:results-subsec:SIA}

\begin{figure}
    \centering
    \includegraphics[width=0.48\textwidth]{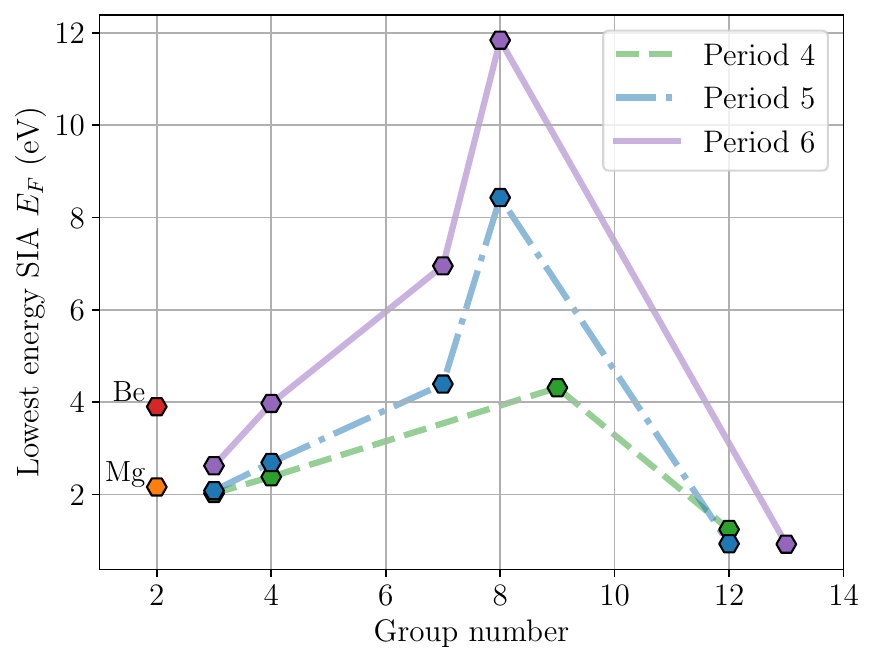}
    \caption{The lowest self-interstitial atom defect formation energy in hcp elements with respect to the group number in the Periodic Table.}
    \label{§3-fig:SIA_Ef_min}
\end{figure}

In figure \ref{§3-fig:SIA_formation_energy} we tabulate the formation energies of SIA configurations in hcp metals across the Periodic Table. %
The lowest formation energy SIA $ E^F_{MIN}(\mathrm{SIA}) $ is highlighted in bold face and white border, and the cells are coloured according to the positive energy difference ${E^F(\mathrm{SIA}) - E^F_{MIN}(\mathrm{SIA}) }$. %
SIA formation energies tend to increase towards group VIII, as shown in figure \ref{§3-fig:SIA_Ef_min} where the minimum formation energy is plotted with respect to the group number. %
Our values for $E_F$ are broadly in agreement with other DFT calculations for Be (\textbf{check}) \cite{Ganchenkova2009, Pasianot2016, Ferry2019}, Mg \cite{Li2015, Pasianot2016}, Y \cite{Borodin2019a}, Ti \cite{Verite2013, Pasianot2016}, Zr \cite{Verite2013}, Hf \cite{Verite2013}, Co \cite{Pasianot2016}, Zn \cite{Pasianot2016,Nitol2021} and Cd \cite{Pasianot2016}. %
For half of the metals with less than ideal $c/a$ ratio, the lowest energy SIA configuration is the basal octahedral SIA. %
In other elements, BO is often close to the lowest energy SIA configuration. %
In particular, \unit{ E^F(\mathrm{BO}) - E^F_{MIN}(\mathrm{SIA}) \lesssim  0.1}{eV} for Mg, Ru, Os, Co and Tl. %

\begin{figure*}
    \centering
    \begin{subfigure}{0.45\textwidth}
        \includegraphics[width=\textwidth]{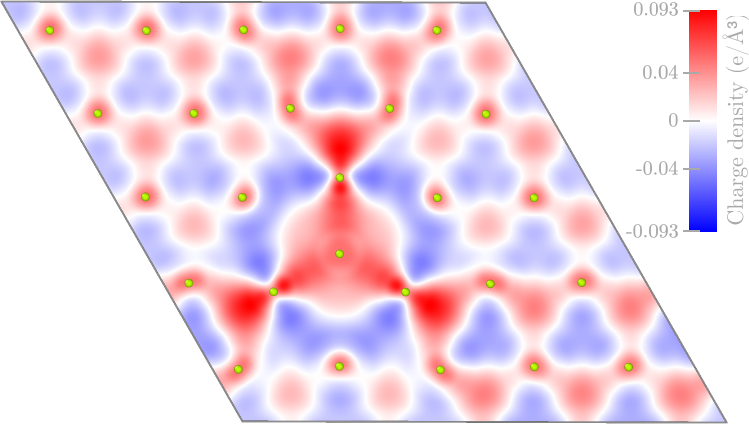}
        \caption{Be}
    \end{subfigure}
    \begin{subfigure}{0.45\textwidth}
        \includegraphics[width=\textwidth]{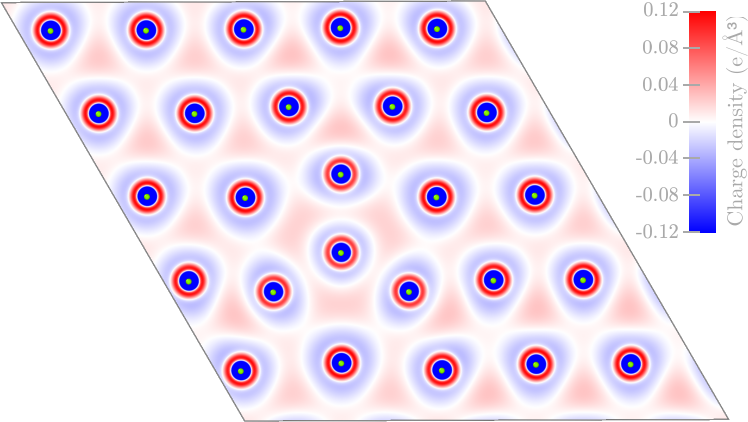}
        \caption{Mg}
    \end{subfigure}
    \begin{subfigure}{0.45\textwidth}
        \includegraphics[width=\textwidth]{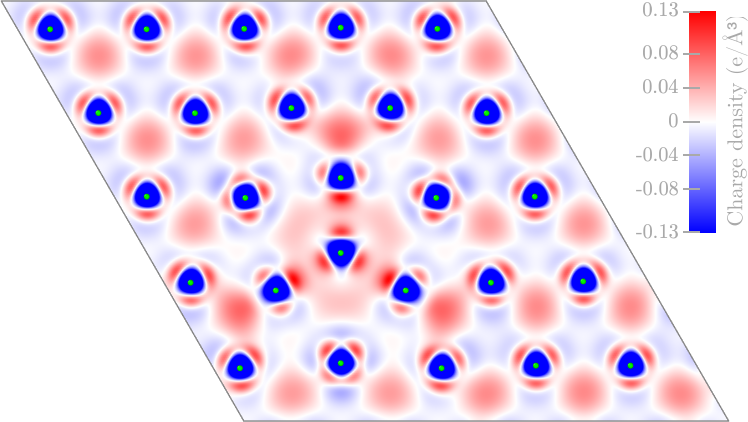}
        \caption{Zr}
    \end{subfigure}
    \begin{subfigure}{0.45\textwidth}
        \includegraphics[width=\textwidth]{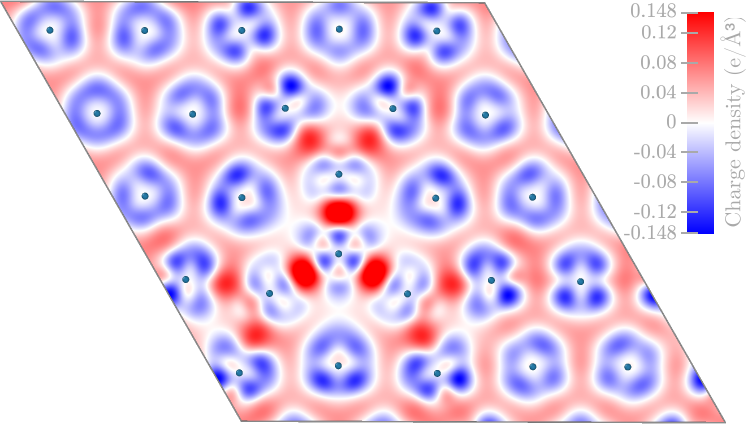}
        \caption{Re}
    \end{subfigure}
    \begin{subfigure}{0.45\textwidth}
        \includegraphics[width=\textwidth]{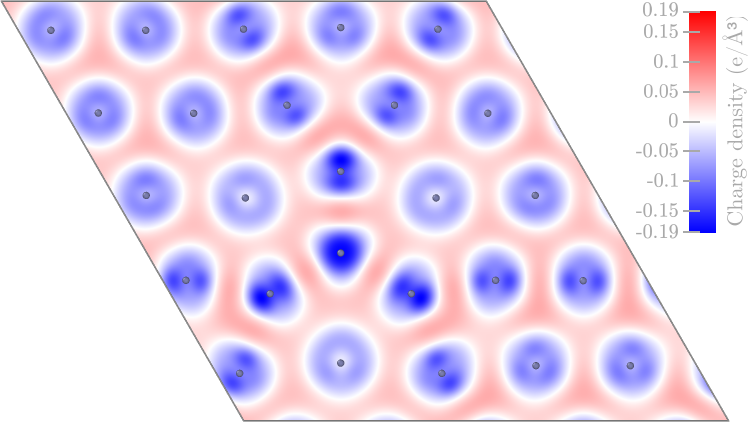}
        \caption{Zn}
    \end{subfigure}
    \begin{subfigure}{0.45\textwidth}
        \includegraphics[width=\textwidth]{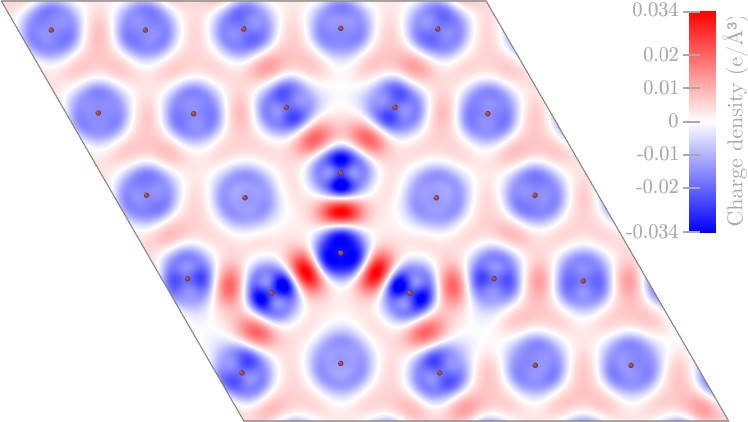}
        \caption{Tl}
    \end{subfigure}
    \caption{ BO SIA defect electron charge density difference (ECDD) for Be, Mg, Zr, Re, Zn and Tl. The position of atoms within the slice are rendered as spheres. }
    \label{§3-fig:BO_ECDD}
\end{figure*}

Crowdion configurations of SIA defects in hcp elements are often unstable and have high formation energies.  %
The linear structure of a crowdion in some bcc metals, like Cr, Mo and W, renders it mechanically unstable against buckling \cite{Ma2019}. %
Such instabilities have also been observed in simulations of crowdion defects in group IV hcp metals Ti, Zr and Hf \cite{Verite2013}. %

The formation energy differences ${E^{\mathrm{BS}}_{F} - E^{\mathrm{BO}}_{F}}$ across group III and IV metals are small. %
\citet{Verite2013} proposed that in group IV elements, SIA defects migrate through a process involving a jump from BO to a buckled crowdion or rotated split dumbbell that then glides along the $c$-axis before hopping to another BO configuration. %
${E^{\mathrm{S}}_{F} - E^{\mathrm{BO}}_{F}}$ is higher for group III elements compared to elements of group IV and thus a BO SIA can migrate within the basal plane by alternating with BS configurations, as suggested for one of the possible migration pathways in Zr \cite{Samolyuk2014}. %

For the group II metals Sc and Y, and the lanthanide Lu, resistivity recovery experiments have shown the onset of stage $I_D$ recovery to be at temperatures $T_{1D}$ between 105 to \unit{130}{K} \cite[§2.4]{Ullmaier1991}. %
This corresponds to migration energies of \unit{\sim 0.3}{eV} and, within the uncertainties of the simulation and experiment, it is not clear whether this is consistent with the lowest formation energy difference of \unit{\sim 0.4 - 0.5}{eV} found in our simulation. %

Our calculations suggest that the most stable SIA in Mg adopts the crowdion configuration. %
However, the difference between the formation energy of a crowdion and any other SIA configuration is less than \unit{0.2}{eV}. %
Therefore, it is difficult to determine the lowest energy SIA with complete certainty, and verify predictions against experimental information. %
Otherwise, we observe that the similarity of formation energies implies that the hcp crystal lattice of Mg is essentially indifferent to the configuration adopted by an extra atom. %
Furthermore, this also implies that the SIA migration is isotropic. %
A similar picture of SIA migration emerges in thallium where, with the exception of a crowdion, the formation energies of all the SIA configurations are the same within \unit{0.12}{eV}. %

Tc and Re in group VII have lower than ideal $c/a$ ratio and yet, despite this, the basal octahedral SIA has a large formation energy. %
The high energy BO, BS and O configurations, together with an unstable crowdion structure imply that SIAs migrate through the rotation of low symmetry dumbbells or buckled crowdions as described by \citet{Verite2013}. %
Understandably, there is little experimental informaton available on the radioactive element Tc. %
For the remaining group VII hcp metal, Re, the SIA migration enthalpy of \unit{0.16}{eV} has been measured experimentally \cite{Ullmaier1991}. %
This differs significantly from our lowest value ${E^F(\mathrm{SIA}) - E^F_{MIN}(\mathrm{SIA}) }$ value of \unit{0.89}{eV}. %
The existence of a lower symmetry rotated dumbbell that has the formation energy close to that of $S$ is possible, or the observed low migration energy may be associated with impurities and imperfections in the Re sample. %

\begin{figure}
    \centering
    \begin{subfigure}{0.45\textwidth}
        \includegraphics[width=\textwidth]{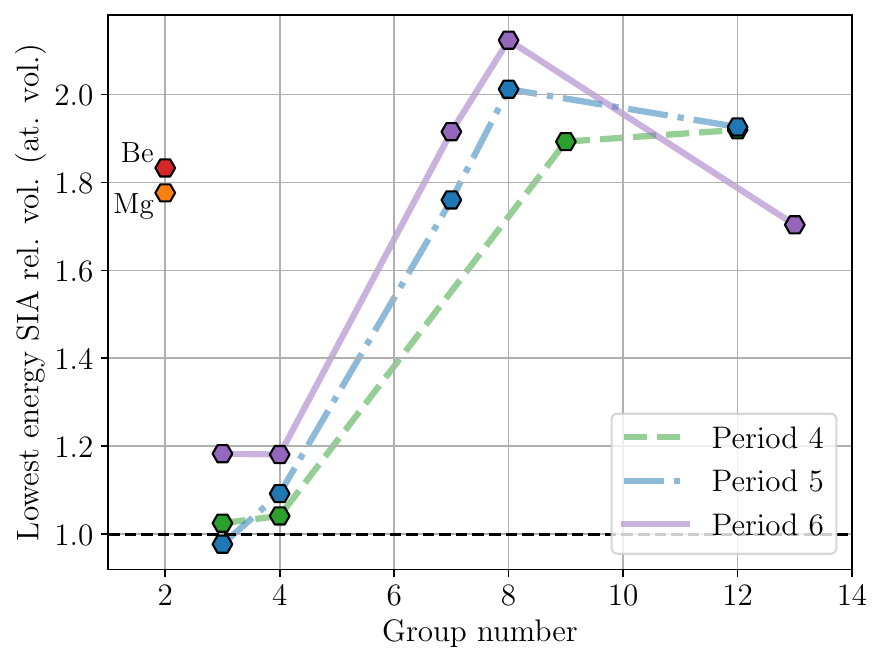}
        \caption{}
    \end{subfigure}
    \begin{subfigure}{0.45\textwidth}
        \includegraphics[width=\textwidth]{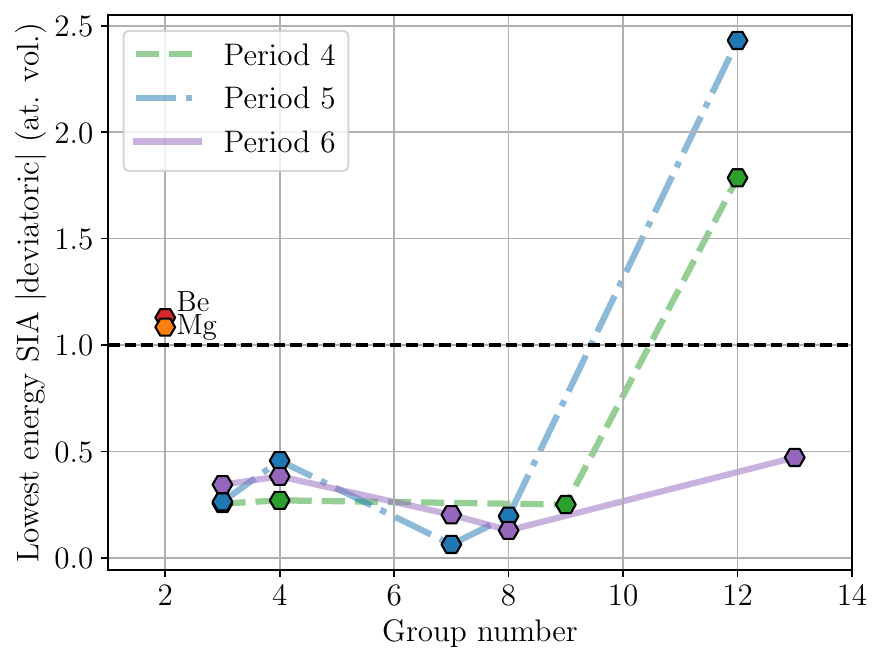}
        \caption{}
    \end{subfigure}
    \caption{Relaxation volumes and the magnitude of relaxation volume deviatoric tensor for the lowest energy SIA defect for each element shown in figure \ref{§3-fig:SIA_formation_energy}.}
    \label{§3-fig:SIA_rel_vols}
\end{figure}

To the best of our knowledge, no experimental information on point defect properties in elements of group VIII is available. %
Our data suggest that $E_f(BS)$ has a high formation energy relative to the lowest energy configurations. %
This raises the possibility that SIAs transform between BO, O and S configurations with relative ease, biasing SIA migration along the $c$ axis. %

Magnetism in Co does not appear to produce any anomalies in the ordering of SIA formation energies. %
Given the lower than ideal $c/a$ ratio of cobalt, the formation energy of the BO configuration is still low. %

Zn and Cd exhibit relatively high formation energies for the basal plane SIAs, whereas the $E_F$ values for O, S and C configurations are similar. %
This implies that the diffusion of SIAs occurs rapidly along the $c$-axis and relatively slow in the basal plane. %
There is a great deal of uncertainty in modelling Zn and Cd using DFT as many exchange-correlation functionals are unable to describe correlations in the $d$-electron shell with sufficient accuracy \cite{Gaston2009}. %
Furthermore, we found that even the predicted lattice parameters and elastic constants of Zn and Cd are highly sensitive to the $k$-point mesh employed. %
We repeated our calculations for Cd with the PBEsol $E_{xc}$ functional \cite{Perdew2008} and found that the formation energy of a crowdion defect was markedly different, with the formation energy of \unit{2.75}{eV} and the lowest formation energy configuration of the defect instead being an octahedral SIA. %
However, the energy of the basal configuration remained significantly higher than that for the inter-basal plane SIAs. %

In figure \ref{§3-fig:BO_ECDD} we show $(0001)$ cross-sections of electron charge density difference (ECDD) for the basal octahedral SIAs in Be, Mg, Zr, Re, and Zn. %
Despite belonging to the same group, Be and Mg exhibit very different bonding characters that are highly anisotropic and isotropic respectively.  %
The $d$-orbital bonding is apparent for Zr and Re whereas the ellipsoidal regions of charge depletion in Zn arises from the completed $d$ shell. %
While the ECDD maps are consistent with the expected bonding in these metals, they do not immediately provide an explanation for the relative stability of different SIA configurations. We note that a similar lack of direct correspondence between the type of interatomic bonding and the structure of defects was found in bcc metals \cite{Ma2019b}. %

The relaxation volumes and magnitude of deviatoric tensors for the lowest formation energy interstitials computed for various chemical elements are plotted in figure \ref{§3-fig:SIA_rel_vols}. %
In a given group, the relaxation volume increases from period 4 to period 6 with the exception of elements of group III. %
Across a given period, the relaxation volume peaks at group VIII. %
With the exception of group XII, the magnitudes of deviatoric tensors across the hcp transition metals are similar. %

\begin{figure}
    \centering
    \includegraphics[width=0.48\textwidth]{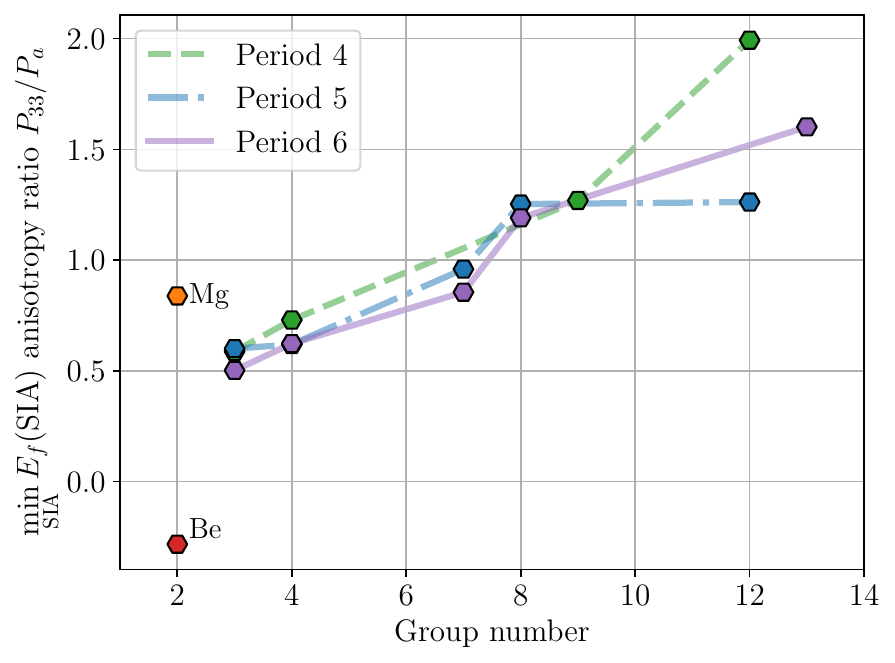}
    \caption{Plot of dipole tensor anisotropy ratio for the lowest energy SIA as a function of the Periodic Table group number for a given element. }
    \label{§3-fig:SIA_lowest_Ef_dipole_anisotropy_ratio}
\end{figure}

Huang diffuse X-ray scattering experiments provide information about the invariants of dipole tensors of low energy SIAs. %
In particular, the anisotropy ratio $P_{33}/P_{a}$ is often reported for hcp metals \cite{Ehrhart1979, Pasianot2016} where it is defined as 
\begin{align}
    P_{a} = \sqrt{\left[2(P^2_{11} + P^2_{22} + 2P^2_{12}) + (P_{11}+P_{22})^2\right]/8}.
\end{align}
Values of parameter $P_a$ for the lowest formation energy SIA configurations are plotted in figure \ref{§3-fig:SIA_lowest_Ef_dipole_anisotropy_ratio} as a function of the group number. %
There is a positive linear correlation along the $d$-block. %
However, we caution that the anisotropy ratio varies strongly as a function of defect configuration and thus, particularly where there is uncertainty as to the true lowest energy SIA, different exchange correlation functionals may produce qualitatively different trends. %
No discernable trend could be identified with respect to the $c/a$ ratio of a given metal. %

\section{HCP, BCC and FCC metals}

\begin{figure}
    \centering
    \begin{subfigure}{0.45\textwidth}
        \includegraphics[width=\textwidth]{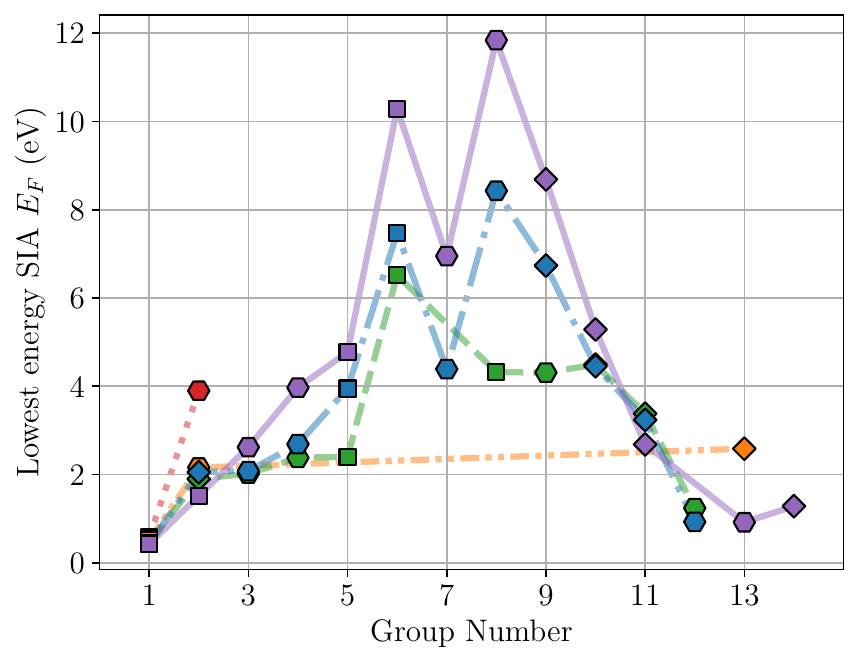}
        \caption{}
    \end{subfigure}
    \begin{subfigure}{0.45\textwidth}
        \includegraphics[width=\textwidth]{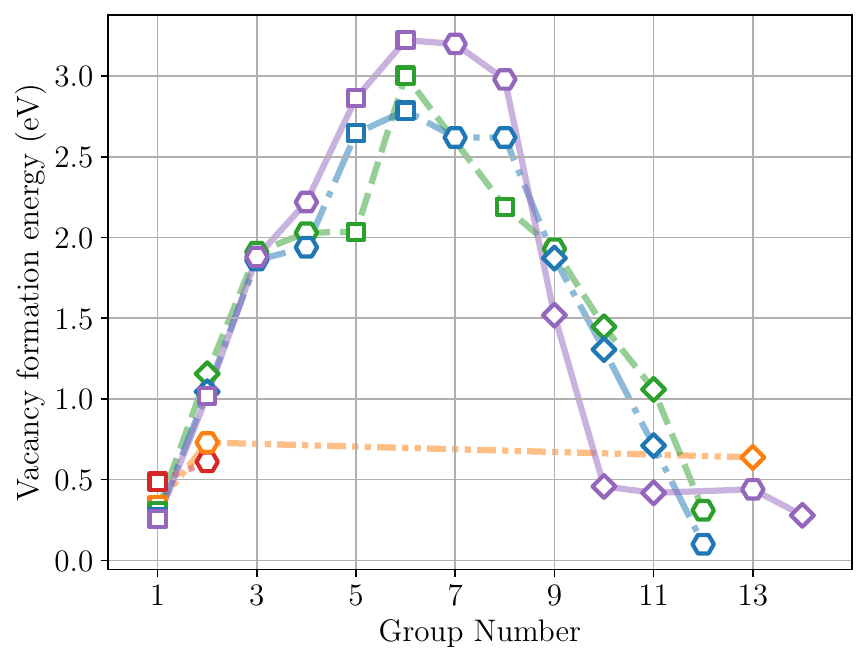}
        \caption{}
    \end{subfigure}
    \caption{Point defect formation energies for SIAs (a) and vacancies (b) across all the bcc, fcc and hcp metals.}
    \label{§4-fig:SIA_Vac_Ef}
\end{figure}

All the new data on point defect relaxation volumes computed for hcp metals have now been combined with the data on bcc \cite{Ma2019b} and fcc \cite{Ma2021} data to provide a comprehensive review of properties of defects across the Periodic Table. %
We compare results obtained using the PBE functional across all the datasets for consistency, and owing to the fact that it was predominantly the PBE functional that was employed across all the elements. %

The formation energies of vacancies and lowest energy SIAs are plotted in figure \ref{§4-fig:SIA_Vac_Ef}. %
The formation energy $E_F$ increases towards the middle of the transition metal block for both types of point defects, with slight variation in the location of the peak between the periods and point defect types. %
Across the third period, vacancy and SIA formation energies are maximum at antiferromagnetic Cr in group VI. %
The largest vacancy formation energy is also at group VI for Mo and W in periods 4 and 5, respectively. %
For the self-interstitial atom defects, $E_F(\mathrm{SIA})$ instead drops substantially at group VII for Tc and Re, before rising slightly at Ru and Os in group VIII. %

\begin{figure}
    \centering
    \includegraphics[width=0.45\textwidth]{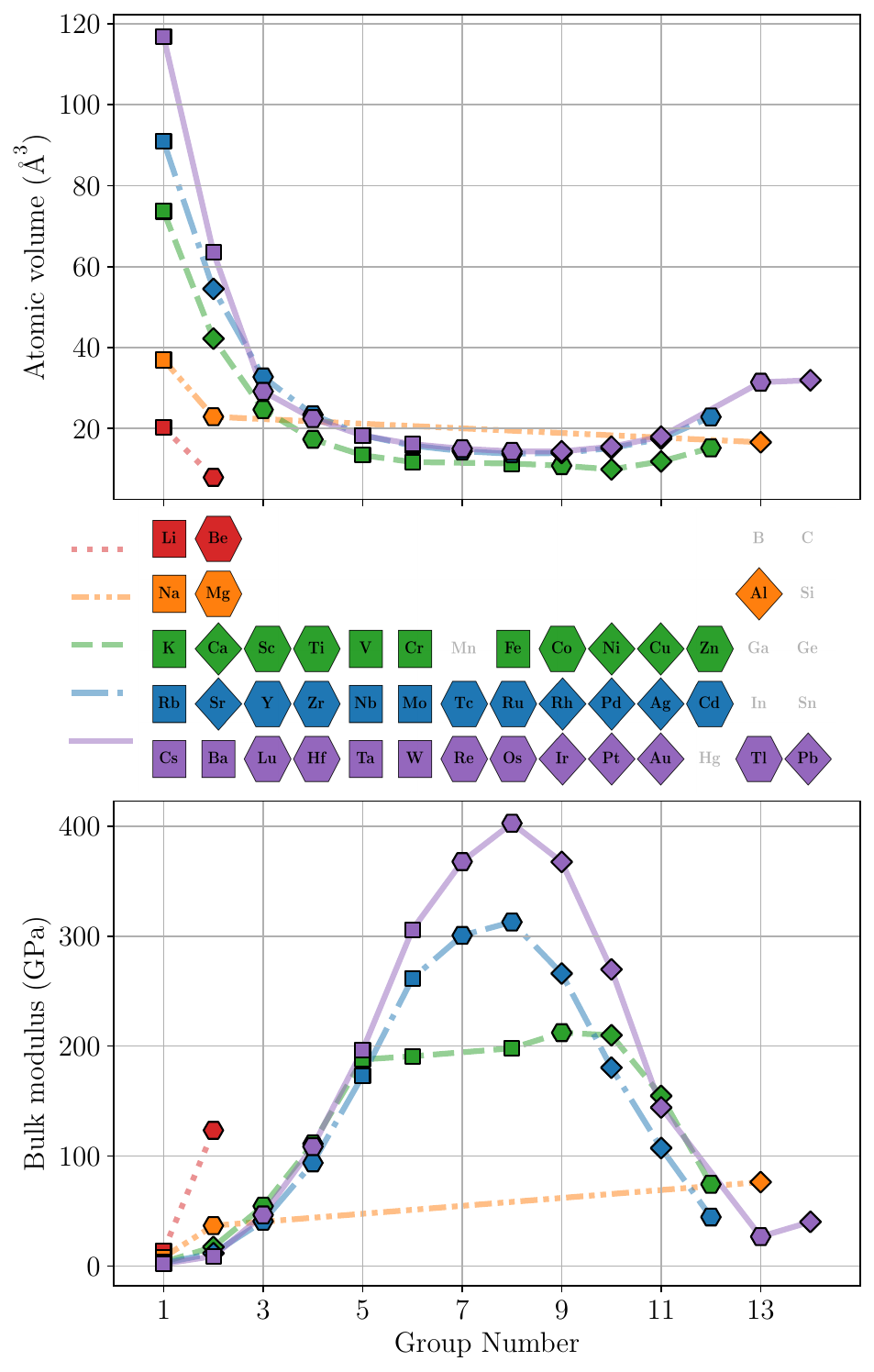}
    \caption{Variation of atomic volume (top) and bulk modulus (bottom) as a function of group number in the Periodic Table for all the bcc, fcc and hcp elements. }
    \label{§4-fig:all_atomic_volume_bulk_moduli}
\end{figure}

These trends with respect to changes in the electronic structure of metals are somewhat similar to the variation of their bulk moduli. %
In particular, figure \ref{§4-fig:all_atomic_volume_bulk_moduli} shows that metals become stiffer towards the middle of the transition block. %
Conversely, the atomic volume is also minimum at that point. %

\begin{figure}
    \centering
    \includegraphics[width=0.45\textwidth]{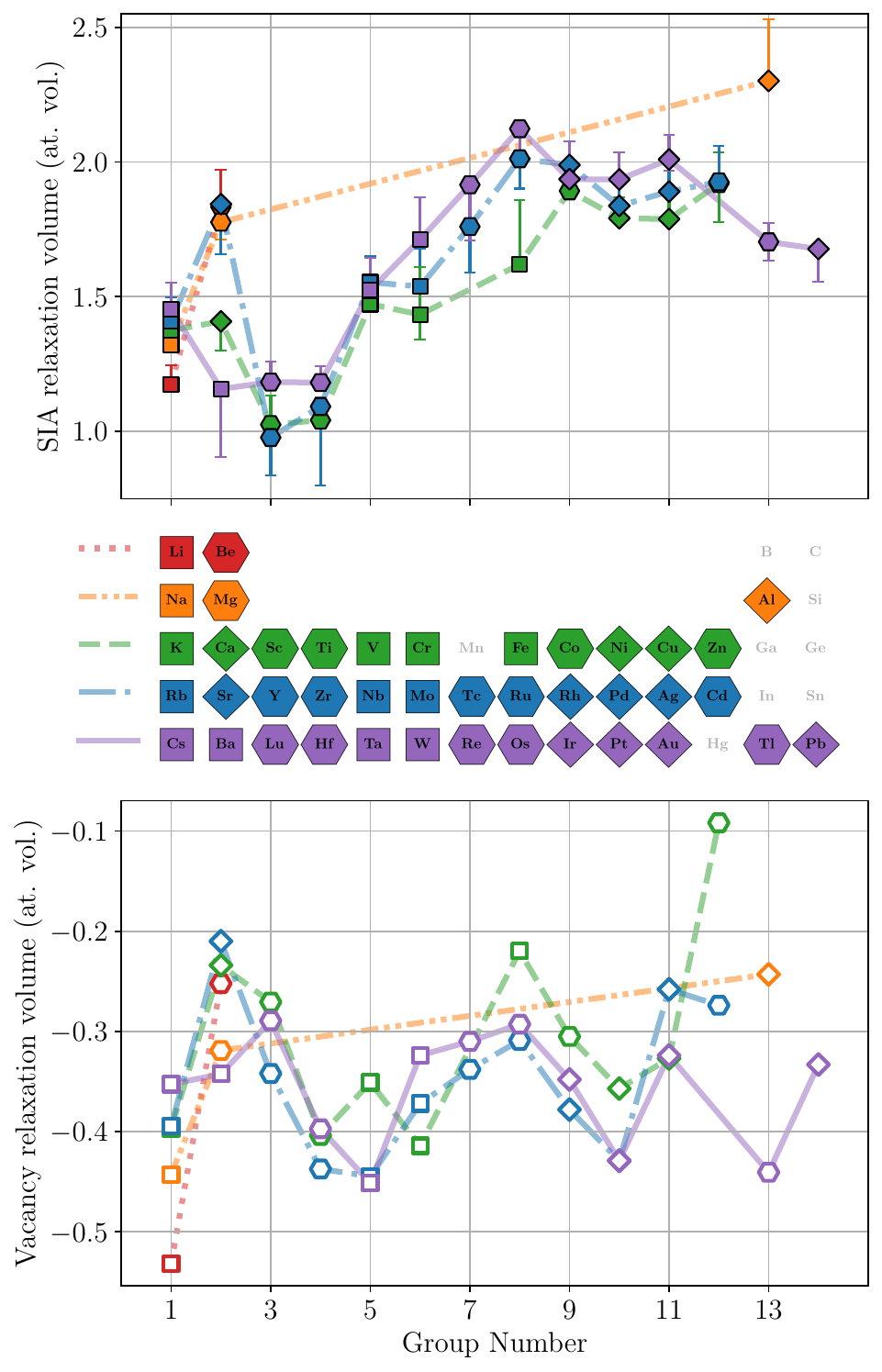}
    \caption{Vacancy relaxation volumes for hcp (this work), bcc \cite{Ma2019b} and fcc \cite{Ma2021} metals. Data for fcc Th have been excluded and hcp Lu was treated as a group 3 element. Error bars indicate the largest and smallest SIA relaxation volume for all the simulated configurations.}
    \label{§3-fig:all_vacancy_SIA_relaxation_volumes}
\end{figure}

In figure \ref{§3-fig:all_vacancy_SIA_relaxation_volumes}, vacancy and SIA relaxation volumes are plotted as functions of element's group number. %
On average, vacancy and self-interstitial atom defect relaxation volumes in pure elements fluctuate around ${\sim -0.35}$ and ${\sim 1.46}$ atomic volumes, respectively. For comparison, in the limit where defects form clusters, for example voids or dislocation loops, the relaxation volume {\it per defect} asymptotically approaches $-1$ atomic volume for vacancy dislocation loops, $+1$ atomic volume for interstitial dislocation loops, and zero for voids \cite{Mason2019}.  %

Variation of relaxation volumes of point defects across a given period is similar to other periods. %
Oscillatory behaviour is observed for vacancies with distinct peaks and troughs in group II, V and VIII with the exception of bcc vanadium. %
Self-interstitial atom defects also show a similar variation across periods, although the oscillation period is larger, with a distinct peak and trough at groups VIII and III, respectively. %

\begin{figure}
    \centering
    \includegraphics[width=0.45\textwidth]{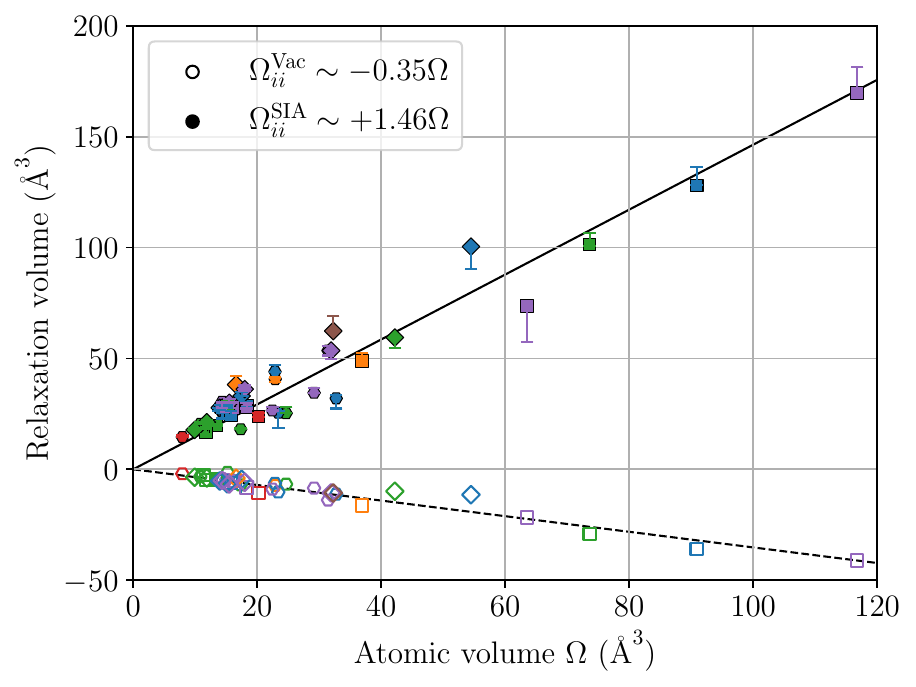}
    \caption{Traces of SIA and vacancy relaxation volume tensors $\Omega _{ii}$ for hcp (this work), bcc \cite{Ma2019b} and fcc \cite{Ma2021} metals plotted against the atomic volume for the respective element. Data points show $\Omega_{ii}$ for the lowest energy SIA configurations, and the error bars indicate the range of variation of defect volumes across all the SIA configurations for that element.}
    \label{§3-fig:SIA_Vac_rel_vols}
\end{figure}

Figure \ref{§3-fig:SIA_Vac_rel_vols} shows a strong correlation between the relaxation volume of a defect and the atomic volume of the respective element. %
The relaxation volumes of vacancies $\Omega_{\mathrm{Vac}}$ and the most stable SIAs $\Omega_{\mathrm{SIA}}$ are well approximated by
\begin{align}
    \Omega_{\mathrm{Vac}} &\sim  -0.35 \Omega \\
    \Omega_{\mathrm{SIA}} &\sim  +1.46 \Omega,
\label{§4-eq:relaxation_volume_fit}
\end{align}
where $\Omega$ is the atomic volume. %
The coefficient of determination $R^2$ for both linear fits is 0.93. %

\begin{figure}
    \centering
    \includegraphics[width=0.45\textwidth]{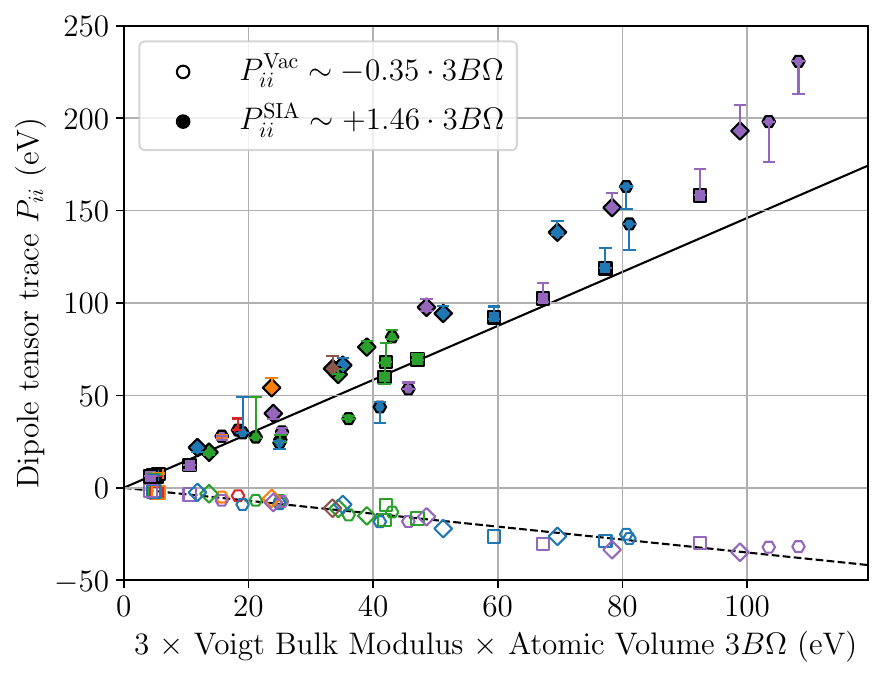}
    \caption{Traces of SIA and vacancy elastic dipole tensors $P_{ii}$ for hcp (this work), bcc \cite{Ma2019b} and fcc \cite{Ma2021} metals plotted against the product of the bulk modulus and atomic volume for the respective element. Data points show $P^{\mathrm{SIA}}_{ii}$ for the lowest energy SIA configurations, and the error bars indicate the range of variation of defect dipole tensors across all the SIA configurations for that element.}
    \label{§4-fig:all_metals_SIA_Vac_dipole_trace}
\end{figure}

The Voigt averaged bulk modulus $B$ provides a good approximation for the pressure $p = - \frac{1}{3}\sigma_{ii}$ and stress tensor $\sigma_{ij}$ arising from dilatation $\Delta$ such that
\begin{equation}
    p = B \Delta,
\label{§4-eq:bulk_modulus}
\end{equation}
where equation \ref{§4-eq:bulk_modulus} is exact for cubic metals. %
Therefore, from equation \ref{§4-eq:relaxation_volume_fit} it follows that the trace of the dipole tensor for vacancies, $P^{\mathrm{Vac}}_{ii}$, and SIAs, $P^{\mathrm{SIA}}_{ii}$ is expected to scale linearly with bulk modulus according to
\begin{align}
    P^{\mathrm{Vac}}_{ii} &\sim  -0.35 \cdot 3 B \Omega \\
    P^{\mathrm{SIA}}_{ii} &\sim  +1.46 \cdot 3 B \Omega,
\end{align}
assuming that the linear regression deduced in equation \ref{§4-eq:relaxation_volume_fit} is satisfied. %
The correlation between the trace of elastic dipole tensor of a defect and the product of the bulk modulus and atomic volume is illustrated in figure \ref{§4-fig:all_metals_SIA_Vac_dipole_trace}. %
The coefficient of determination $R^2$ for SIA and vacancies dipole tensor trace linear fits are 0.86 and 0.93, respectively. %
The gradient of 1.79 presents a better fit for the SIA dipole tensor trace, yielding $R^2 = 0.94$. %

No notable trends were identified for the anisotropic parts of the dipole and relaxation volume tensors with respect to electron configuration. %
This suggests that the anisotropic properties of point defects in metals is most strongly affected by the crystal structure. %
For example, bcc and fcc vacancies contain no deviatoric components in either their dipole or relaxation volume tensors \cite{Ma2019b,Ma2021}. %
However, in hcp metals the deviatoric part can be substantial as discussed in section \ref{sec:results-subsec:vacancies}. %

 \section{Conclusions}
\label{sec:conclusions}

In this study, we computed elastic dipole and relaxation volume tensors of intrinsic point defects across all the hcp metals in the periodic table. %
This completes the {\it ab initio} database of point defect properties of hcp, bcc and fcc metals \cite{Ma2019b, Ma2021} across the Periodic Table, save the lanthanides and actinides. %
SIA defects show little variation in their relaxation volumes for the configurations considered; it remains to be seen whether lower symmetry configurations break this trend. %
For both interstitials and vacancies, we have found a strong correlation between the traces of relaxation volume tensors and dipole tensors with atomic volume and bulk modulus of respective elements. %
The data reviewed in this study can now be used for predicting radiation-induced deformation \cite{Reali2022} of metals both currently in service in nuclear power plants and those yet to be tested in  extreme radiation environments. %

\foreach \Element/\Eexc/\E/\exc in {Beryllium/Be/Be/PBE,Beryllium/Be_PBEsol/Be/PBEsol,Magnesium/Mg/Mg/PBE,Scandium/Sc/Sc/PBE,Yttrium/Y/Y/PBE,Lutetium/Lu/Lu/PBE,Titanium/Ti/Ti/PBE,Zirconium/Zr/Zr/PBE,Hafnium/Hf/Hf/PBE,Technetium/Tc/Tc/PBE,Rhenium/Re/Re/PBE,Ruthenium/Ru/Ru/PBE,Osmium/Os/Os/PBE,Cobalt/Co/Co/PBE,Zinc/Zn/Zn/PBE,Zinc/Zn_PBEsol/Zn/PBEsol,Cadmium/Cd/Cd/PBE,Cadmium/Cd_PBEsol/Cd/PBEsol,Thallium/Tl/Tl/PBE}{
    \DefectDataTable{\Element}{\Eexc}{\E}{\exc}
}

\begin{acknowledgments}
The authors gratefully acknowledge stimulating discussions with M. Boleininger and D. R. Mason. %
This work has been carried out within the framework of the EUROfusion Consortium, funded by the European Union via the Euratom Research and Training Programme (Grant Agreement No. 101052200—EUROfusion), and was partially supported by the Broader Approach Phase II agreement under the PA of IFERC2-T2PA02. %
This work was also funded by the UK Fusion Futures Programme and the UK EPSRC Energy Programme (Grant Number EP/W006839/1). %
To obtain further information on the data and models underlying the paper please contact PublicationsManager@ukaea.uk. %
Views and opinions expressed are however those of the authors only and do not necessarily reflect those of the European Union or the European Commission. %
Neither the European Union nor the European Commission can be held responsible for them. %
Numerical calculations were performed using resources provided by the Cambridge Service for Data Driven Discovery (CSD3) operated by the University of Cambridge Research Computing Service and ARCHER 2, the UK National Supercomputing Service. %
\end{acknowledgments}

\appendix

\section{Convergence testing}

Here we detail convergence tests for all of the hexagonal close packed (hcp) metals discussed in the main text. Each figure from figure 19 to 37 shows the results of convergence tests and elastic dipole tensor extrapolations for a given element using the Perdew-Burke-Ernzerhof (PBE) exchange correlation functional. Where the PBEsol was considered this is stated after the element name for Be, Cd and Zn in separate figures. 

For each element, the plane wave energy cut-off $E_{\mathrm{cut}}$ and $k$-point mesh data are shown in subfigures (a) and (b). The convergence tests were conducted as follows. A primitive hcp unit cell containing two atoms had the atom $(1/3,2/3,1/4)$ displaced by ${0.1 \, \text{\r{A}}}$ along the $x$ direction. %
The $k$-point mesh was fixed as $E_{\mathrm{cut}}$ was varied in subfigure (a) and $E_{\mathrm{cut}}$ was fixed as the $k$-point mesh was varied. $k$ meshes ${k_a \times k_a \times k_c }$ were $\Gamma$-centred and chosen such that the ratio of grid points along the basal plane reciprocal lattice vectors $k_a$ to the number of points along the $c$ direction $k_c$ was approximately equal to the experimental $c/a$ ratio of the given hcp metal. A single self-consistent field (SCF) cycle was subsequently carried out. The energy per atom (eV), $[2 \bar 1 \bar 1 0]$ component of force on the displaced atom (eV/\AA) and the $11$ component of the unit cell stress were recorded for each SCF calculation. For all elements, we attempted to choose $E_{\mathrm{cut}}$ and the $k$-point mesh so that the energy per atom, force and stress quantities were converged to at least 1 meV per atom, 1 meV/\AA\ and 0.5 kBar respectively. Where this could not be achieved, we subsequently chose the highest plane wave cut-off and/or densest $k$-point mesh tested.

Subfigures (c) and above show the dipole tensor calculations for vacancy and self-interstitial atom defects. Where a SIA was unstable and did not relax satisfactorily the corresponding plots are not included. 

\foreach \Element/\E in {Beryllium/Be,Beryllium PBEsol/Be_PBEsol,Magnesium/Mg,Scandium/Sc,Yttrium/Y,Lutetium/Lu,Titanium/Ti,Zirconium/Zr,Hafnium/Hf,Technetium/Tc,Rhenium/Re,Ruthenium/Ru,Osmium/Os,Cobalt/Co,Zinc/Zn,Zinc PBEsol/Zn_PBEsol,Cadmium/Cd,Cadmium PBEsol/Cd_PBEsol,Thallium/Tl} {
    \begin{figure*}[t]
    \centering
    \begin{subfigure}[t]{0.45\textwidth}
        \centering
        \includegraphics[width=\textwidth]{Supp_figures/\E/ENCUT.pdf}
        \caption{$E_{cut}$ convergence test}
    \end{subfigure}%
    ~ 
    \begin{subfigure}[t]{0.45\textwidth}
        \centering
        \includegraphics[width=\textwidth]{Supp_figures/\E/KPOINTS.pdf}
        \caption{$k$-mesh convergence test}
    \end{subfigure}

    \begin{subfigure}[t]{0.4\textwidth}
        \centering
        \includegraphics[width=\textwidth]{Supp_figures/\E/Vac_dipole_tensor.pdf}
        \caption{Vacancy $P_{ij}$}
    \end{subfigure}%
    ~ 
    \begin{subfigure}[t]{0.4\textwidth}
        \centering
        \includegraphics[width=\textwidth]{Supp_figures/\E/O-a_dipole_tensor.pdf}
        \caption{Octahedral SIA $P_{ij}$}
    \end{subfigure}

    \begin{subfigure}[t]{0.4\textwidth}
        \centering
        \includegraphics[width=\textwidth]{Supp_figures/\E/BO-b_dipole_tensor.pdf}
        \caption{Basal octahedral SIA $P_{ij}$}
    \end{subfigure}%
    ~ 
    \begin{subfigure}[t]{0.4\textwidth}
        \centering
        \includegraphics[width=\textwidth]{Supp_figures/\E/S-2f_dipole_tensor.pdf}
        \caption{$\langle 0001 \rangle$ split dumbbell SIA $P_{ij}$}
    \end{subfigure}

    \IfFileExists{Supp_figures/\E/BS-2j_dipole_tensor.pdf}{
    \begin{subfigure}[t]{0.4\textwidth}
        \centering
        \includegraphics[width=\textwidth]{Supp_figures/\E/BS-2j_dipole_tensor.pdf}
        \caption{$\langle 2 \bar 1 \bar 1 0\rangle$ basal split dumbbell SIA $P_{ij}$}
    \end{subfigure}%
    }
    ~ 
    \IfFileExists{Supp_figures/\E/C-g_dipole_tensor.pdf}{
    \begin{subfigure}[t]{0.4\textwidth}
        \centering
        \includegraphics[width=\textwidth]{Supp_figures/\E/C-g_dipole_tensor.pdf}
        \caption{Crowdion SIA $P_{ij}$}
    \end{subfigure}
    }
    
    \caption{\Element}
\end{figure*}
}

\clearpage 
\bibliography{references}

\end{document}